\documentclass[11pt]{article}

\usepackage[T1]{fontenc}
\usepackage[utf8]{inputenc}
\usepackage{mathptmx}
\usepackage[margin=1in]{geometry}
\usepackage{microtype}
\usepackage{xcolor}
\usepackage{amsmath,amssymb,amsthm,mathtools,bm}

\usepackage{booktabs,array,graphicx}
\usepackage[labelsep=period]{caption}
\usepackage{enumitem}

\usepackage[numbers,sort&compress]{natbib}
\usepackage[hidelinks]{hyperref}
\hypersetup{pdftitle={Random weighting inference for linear regression under serial dependence of unknown form},
            pdfauthor={Hamadou, Longla, Muia, Hasan}}

\theoremstyle{plain}
\newtheorem{theorem}{Theorem}[section]
\newtheorem{proposition}{Proposition}[section]
\newtheorem{corollary}{Corollary}[section]
\newtheorem{lemma}{Lemma}[section]
\theoremstyle{definition}
\newtheorem{definition}{Definition}[section]

\theoremstyle{remark}

\makeatletter
\renewenvironment{proof}[1][\proofname]{\par
  \pushQED{\qed}%
  \normalfont \topsep12\p@\@plus4\p@\relax
  \trivlist
  \item[\hskip\labelsep\bfseries #1\@addpunct{.}]\ignorespaces
}{%
  \popQED\endtrivlist\@endpefalse
}
\makeatother
\numberwithin{equation}{section}

\newcommand{\E}{\mathbb{E}}
\newcommand{\Prob}{\mathbb{P}}
\DeclareMathOperator{\Var}{Var}
\DeclareMathOperator{\Cov}{Cov}
\DeclareMathOperator{\tr}{tr}

\newcommand{\R}{\mathbb{R}}
\newcommand{\T}{\top}
\newcommand{\ind}{\mathbf{1}}
\newcommand{\dto}{\Rightarrow}
\newcommand{\pto}{\xrightarrow{\;\Prob\;}}

\newcommand{\zt}{z_t}
\newcommand{\et}{\varepsilon_t}
\newcommand{\st}{s_t}
\newcommand{\sbar}{\bar{s}_n}
\newcommand{\Qm}{Q}
\newcommand{\Vm}{V}
\newcommand{\Wm}{W}
\newcommand{\Ln}{L_n}
\newcommand{\Lm}{L}
\newcommand{\wnt}{w_{nt}}
\newcommand{\taun}{\tau_n}
\newcommand{\neff}{n_{\mathrm{eff}}}
\newcommand{\rhostar}{\rho^{\ast}}
\newcommand{\kbar}{\bar{\kappa}}
\newcommand{\Gn}{G_n}
\newcommand{\bhat}{\hat{\beta}^{w}_{n}}
\newcommand{\bols}{\hat{\beta}_{\mathrm{ols}}}
\newcommand{\Chat}{\hat{C}_n}
\newcommand{\ehat}{\hat{e}_t}
\newcommand{\xit}{\xi_t}
\newcommand{\hn}{h_n}

\title{Random weighting inference for linear regression under serial dependence of unknown form}

\author{%
Mous-Abou Hamadou\thanks{Department of Mathematics and Statistics, Quinnipiac University, Hamden, CT 06518, USA.
  Corresponding author, mhamadou@quinnipiac.edu}
\and
Martial Longla\thanks{Department of Mathematics, The University of Mississippi, University, MS 38677, USA}
\and
Mathias Nthiani Muia\thanks{Department of Mathematics and Statistics, University of South Alabama, Mobile, AL 36688, USA}
\and
Mahmud Hasan\thanks{Department of Biostatistics, Virginia Commonwealth University, Richmond, VA 23219, USA}}

\date{}

\begin{document}
\maketitle

\begin{abstract}
We construct confidence regions for a linear regression's coefficients, with regressors and errors forming a stationary ergodic sequence of serial dependence not necessarily known. We weight the observations by random variables from an independent auxiliary sample and estimate the coefficients by weighted least squares. We prove the estimator to be asymptotically normal with the sandwich covariance if and only if the bandwidth shrinks fast enough for the weights' randomness to dominate the score's serial dependence. Wald regions have the nominal level, under short memory with the dependence left unspecified and under long memory with a non-normal least-squares limit. We derive a condition specific to the slopes and a bandwidth from consistent memory-order estimates, and identify the estimator asymptotically with least squares perturbed by independent noise. In simulations, we compare our regions with self-normalized, fixed-bandwidth, and resampled intervals under short and long memory, and we apply the method to daily wind speeds at Irish stations, a long-memory series.
\end{abstract}



\section{Introduction}\label{sec:intro}

In regression on time series, confidence regions for the coefficients depend on the long-run covariance of the regression score, the sum of the score's autocovariances at every lag. Under dependence conditions implying a central limit theorem for the score mean, \citet{NeweyWest1987} and \citet{Andrews1991}, among others, estimate the long-run covariance by smoothing the sample autocovariances with a kernel. Along the same lines, \citet{KieferVogelsang2002}, for example, fix the ratio of the bandwidth to the sample size and adjust the critical values, and \citet{Shao2010} divides by a normalizer from recursive coefficient estimates. Others resample blocks of observations or dependent multipliers to reproduce the dependence, for instance \citet{Kunsch1989}, \citet{PolitisRomano1994}, and \citet{Shao2010DWB}.

Stationarity and ergodicity yield the law of large numbers for the score mean, yet \citet{Bradley1989} and \citet{Volny1990} constructed stationary sequences with finite variance violating the central limit theorem. Consistent estimation of the long-run variance from a single sample depends in turn on restrictions on the dependence (see \citet{KojevnikovSong2023}). Under long memory, the long-run variance is infinite, and for the mean \citet{McElroyPolitis2012} found a fixed-bandwidth limit of a different form. Analyses of bandwidth choice, including \citet{BaillieDieboldKapetaniosKimMora2025}, exclude long memory by assumption. With long memory in a regressor and in the error, least squares converges slower than $\sqrt n$ to a non-normal limit, for instance in the models of \citet{TsayChung2000}, \citet{Chung2002}, and \citet{GuoKoul2008}. In response, \citet{RobinsonHidalgo1997} recover $\sqrt n$-consistency and asymptotic normality under long memory with weighted least squares. For such data, we look for the normal limit in randomness added by the analyst.

In the case of independent observations, \citet{Rubin1981}, \citet{Zheng1987}, \citet{MasonNewton1992}, and \citet{PraestgaardWellner1993}, among others, studied random weighting, with a weighted statistic's conditional variance given the data approximating the sampling variance. Under dependence, \citet{LongPeli2021} extended random weighting to the mean of a strictly stationary ergodic sequence of finite variance, drawing the weights from an auxiliary sample independent of the data. Given the data, the weighted sum has independent summands and satisfies the Lindeberg condition under every form of serial dependence, the normal limit coming from the weights. Dependence remains in the sample mean, negligible under a condition on the bandwidth, and the weighted mean is then asymptotically normal with a variance proportional to the marginal variance. With weights from random sampling times, \citet{DavydovTempelman2024} proved the same for every stationary ergodic process with finite variance. In regression, \citet{Leung2022} applied resampled statistics under $\sqrt n$-consistency and weak dependence of unknown form, at the cost of a slower rate and a conclusion specific to the auxiliary sample, and \citet{Song2016} had proposed a randomized confidence function for the second.

We extend random weighting to regression under strong dependence, estimating the coefficients by weighted least squares. To first order, the weighted estimator's error is the sum of a term from the weights, normal in the limit, and the least-squares error with the data's serial dependence. Dominance of the former over the latter is a condition on the score mean, and we prove the condition to be necessary and sufficient for a normal limit with the sandwich covariance. For the slopes, a weighted Frisch--Waugh reduction yields a weaker condition when the intercept coordinate has the strongest memory. Under ergodicity, some bandwidth satisfies either condition, and we compute one in practice from consistent memory-order estimates.

Randomization costs a slower rate and a conclusion specific to the auxiliary sample, and we quantify the cost. Intervals and detectable deviations shrink like the inverse square root of the effective sample size, and at a fixed bandwidth, the coverage given the data is a noncentral chi-squared probability. From the formula follow coverage and width under a central limit theorem and the limits of constructions less dependent on the auxiliary sample. We take the variance identity and the conditional Lindeberg argument from \citet{LongPeli2021} and \citet{Leung2022}. To our knowledge, the rate condition's necessity, the slope condition, the conditional coverage formula, and the limits of the constructions less dependent on the auxiliary sample are new.

We define the randomly weighted least-squares estimator and state the variance identity in Section~\ref{sec:method}. In Section~\ref{sec:theory}, we prove asymptotic normality with the sandwich covariance and derive the slope condition and the bandwidth from memory-order estimates. Conditional coverage at a fixed second moment of the weights and the limits of constructions less dependent on the auxiliary sample follow in Section~\ref{sec:cost}. We report the simulations in Section~\ref{sec:evidence} and the application to Irish wind speeds in Section~\ref{sec:application}, and close with a discussion in Section~\ref{sec:discussion} and further proofs in the appendix.

\subsection{Notation}\label{sec:notation}

In what follows, $\|x\|$ is the Euclidean norm of a vector, $\|A\|$ the Frobenius norm of a matrix, $A^\T$ the transpose, $\tr(A)$ the trace, $I_m$ the identity of order $m$, $e_j$ the $j$th unit vector of $\R^m$, $\ind\{\cdot\}$ an indicator, and $\lfloor x\rfloor$ the integer part. As $n\to\infty$, $\dto$ and $\pto$ denote convergence in distribution and in probability, $O_{\Prob}$ and $o_{\Prob}$ the stochastic orders, and $a_n\sim b_n$ means $a_n/b_n\to1$.

\section{Randomly weighted least squares}\label{sec:method}

Let $(X_t,Y_t)_{t\ge1}$ be a sequence with $X_t\in\R^p$ and $Y_t\in\R$, observed for $t=1,\dots,n$, set $\zt=(1,X_t^\T)^\T\in\R^m$ with $m=p+1$, and let
\begin{equation}\label{eq:model}
Y_t=\zt^\T\beta+\et,\qquad \E[\zt\et]=0,\qquad t\ge1,
\end{equation}
for some $\beta\in\R^m$, under the following condition. \begin{description}[leftmargin=2.6em,labelindent=0pt]
\item[(A1)] $(X_t,Y_t)_{t\ge1}$ is strictly stationary, $\E\|z_1\|^2<\infty$, $\E Y_1^2<\infty$, $\E\|z_1\varepsilon_1\|^2<\infty$, and $\Qm=\E[z_1z_1^\T]$ is positive definite.
\end{description}
Under (A1), $\beta=\Qm^{-1}\E[z_1Y_1]$ by \eqref{eq:model}, and the score is $\st=\zt\et$. We define
\[
\sbar=\frac1n\sum_{t=1}^n\st,\qquad \Vm=\E[s_1s_1^\T],\qquad \Ln=n\Var(\sbar)\quad \text{ and }\Wm=\Qm^{-1}\Vm\Qm^{-1}.
\]
With $\hat\Qm_n=\frac1n\sum_{t=1}^n\zt\zt^\T$, the least-squares estimator of $\beta$ is
\[
\bols=\begin{cases}\hat\Qm_n^{-1}\,\dfrac1n\displaystyle\sum_{t=1}^n\zt Y_t & \text{if }\hat\Qm_n\text{ is nonsingular},\\[6pt] 0 & \text{otherwise}.\end{cases}
\] If the scores are serially uncorrelated and their mean satisfies a central limit theorem, then $\sqrt n(\bols-\beta)\dto N(0,\Wm)$, where $\Wm$ is the sandwich covariance of \citet{White1980}. If the autocovariances of the score are summable, $\Ln$ converges to the long-run covariance $\Lm$ of the score, and if the score has long memory, $\Ln$ is unbounded.

We randomize least squares by multiplying the squared residuals by random variables independent of the data.

\begin{definition}\label{def:weights}
A triangular array $(\wnt)_{1\le t\le n,\,n\ge1}$ is a random weighting with second moment $\taun\ge1$ if, for each $n$, $w_{n1},\dots,w_{nn}$ are independent and identically distributed, nonnegative, independent of $(X_t,Y_t)_{t\ge1}$, with $\E\wnt=1$ and $\E\wnt^2=\taun$, and $\neff=n/\taun$ is the effective sample size.
\end{definition}

As $\taun=1+\Var(\wnt)$, ordinary least squares corresponds to $\taun=1$ and $\wnt=1$ for every $t$, and, at a given $n$, a larger $\taun$ makes the weighted fit depend more on the weights and less on the serial dependence of the data. For fixed numbers $a_1,\dots,a_n$ bounded by $C$, the weighted average $n^{-1}\sum_t\wnt a_t$ has mean $\bar a_n=n^{-1}\sum_ta_t$ and variance $(\taun-1)n^{-2}\sum_ta_t^2\le C^2/\neff$, so in the variance of a sample average, the weights act as a reduction of the sample size from $n$ to $\neff$. With $\Gn=\frac1n\sum_{t=1}^n\wnt\zt\zt^\T$, the randomly weighted least-squares estimator is
\begin{equation}\label{eq:estimator}
\bhat=\begin{cases}\Gn^{-1}\,\dfrac1n\displaystyle\sum_{t=1}^n\wnt\zt Y_t & \text{if }\Gn\text{ is nonsingular},\\[6pt] \bols & \text{otherwise}.\end{cases}
\end{equation}

Independent draws from any nonnegative distribution with mean one and finite variance satisfy Definition~\ref{def:weights}. For the mean of a stationary ergodic sequence, \citet{LongPeli2021} used kernel weights to construct confidence intervals. In their construction, $K$ is a bounded probability density on $\R$, the auxiliary variables $\xi_1,\xi_2,\dots$ are independent with a bounded density $f$, continuous at $0$ with $f(0)>0$, and independent of the data, and $\hn>0$ is a bandwidth. Their weights, normalized to mean one, and the second moment are
\begin{equation}\label{eq:kernel}
\wnt=\frac{K(\xit/\hn)}{\E K(\xi_1/\hn)},\qquad \taun=\frac{\E K^2(\xi_1/\hn)}{\{\E K(\xi_1/\hn)\}^2}.
\end{equation}
With $c_j(K)=\int K^j(u)\,du$, the substitution $x=\hn u$ turns $\E K^j(\xi_1/\hn)$ into $\hn\int K^j(u)f(\hn u)\,du$, and, as $\hn\to0$, dominated convergence and the continuity of $f$ at $0$ lead to
\[
\hn\taun\to\frac{c_2(K)}{f(0)},\qquad
\frac{\E\wnt^4}{\taun^3}\to\frac{c_4(K)}{c_2(K)^3},\qquad
\frac{\sup K}{\taun\,\E K(\xi_1/\hn)}\to\frac{\sup K}{c_2(K)} .
\]
Thus $\neff^{1/2}$ is, up to a constant, the rate $(n\hn)^{1/2}$ of \citet{LongPeli2021}. Since $\wnt\le\sup K/\E K(\xi_1/\hn)$, the third limit ensures $\wnt\le c_0\taun$ for a constant $c_0$ and all large $n$. For the Gaussian kernel and standard normal $\xit$, $\taun=(1+\hn^2)/\{\hn(2+\hn^2)^{1/2}\}$, and $c_4(K)/c_2(K)^3=\sqrt2$.

In random subsampling, with independent Bernoulli variables $I_{n1},\dots,I_{nn}$ of success probability $\pi_n\in(0,1]$, independent of the data, retaining observation $t$ when $I_{nt}=1$ corresponds to the weights and second moment
\begin{equation}\label{eq:subsample}
\wnt=\frac{I_{nt}}{\pi_n},\qquad \taun=\frac1{\pi_n},
\end{equation}
with $\E\wnt^4=\taun^3$ and $\wnt\le\taun$. Then $\bhat$ is the least-squares estimator computed from the subsample $\{t\mid I_{nt}=1\}$, of expected size $n\pi_n=\neff$.

We call $1/\taun$ the bandwidth of a random weighting, equal to $\pi_n$ for subsampling weights and asymptotic to $\hn f(0)/c_2(K)$ for kernel weights.

We write $\mathcal D_n$ for the $\sigma$-field generated by $(X_t,Y_t)$, $1\le t\le n$.

\begin{proposition}\label{prop:identity}
Assume \textup{(A1)}, let $(\wnt)$ be a random weighting with second moment $\taun$, and set
\[
A_n=\frac1n\sum_{t=1}^n(\wnt-1)\st .
\]
Then, for every $n$,
\begin{enumerate}[label=\textup{(\roman*)},leftmargin=2.4em]
\item $\bhat-\beta=\Gn^{-1}(A_n+\sbar)$ whenever $\Gn$ is nonsingular,
\item $\E[A_n\mid\mathcal D_n]=0$ and
\[
\Cov(A_n\mid\mathcal D_n)=\frac{\taun-1}{n^2}\sum_{t=1}^n\st\st^\T ,
\]
\item $\Cov(A_n,\sbar)=0$ and
\[
n\Var\Big(\frac1n\sum_{t=1}^n\wnt\st\Big)=(\taun-1)\Vm+\Ln .
\]
\end{enumerate}
\end{proposition}

Proposition~\ref{prop:identity} is proved in the appendix.

\section{Asymptotic normality and confidence regions}\label{sec:theory}

We assume the following laws of large numbers for the design and for the squared scores.
\begin{description}[leftmargin=2.6em,labelindent=0pt]
\item[(A2)] $\hat\Qm_n\pto\Qm$ and $n^{-1}\sum_{t=1}^n\st\st^\T\pto\Vm$.
\end{description}
When $(X_t,Y_t)_{t\ge1}$ is ergodic, (A2) follows from (A1). We choose the weights to satisfy the following condition.
\begin{description}[leftmargin=2.6em,labelindent=0pt]
\item[(A3)] $\taun\to\infty$ and $\E w_{n1}^4/(n\taun^2)\to0$.
\end{description}
By H\"older's inequality,
\[
\E w_{n1}^4\ge(\E w_{n1}^2)^3/(\E w_{n1})^2=\taun^3,
\]
so the second part of (A3) implies $\neff\to\infty$ and, for every $c>0$, the Lindeberg condition
\[
\frac1{\taun}\E\big[(w_{n1}-1)^2\ind\{|w_{n1}-1|>c\,(n\taun)^{1/2}\}\big]\le\frac{\E(w_{n1}-1)^4}{c^2n\taun^2}\to0 .
\]

If $\wnt\le c_0\taun$ for a constant $c_0$ and all large $n$, then $\E w_{n1}^4\le c_0^2\taun^3$, and (A3) reduces to $\taun\to\infty$ and $\neff\to\infty$. Kernel weights \eqref{eq:kernel} therefore satisfy (A3) when $\hn\to0$ and $n\hn\to\infty$, the bandwidth conditions of \citet{LongPeli2021}, and subsampling weights \eqref{eq:subsample} when $\pi_n\to0$ and $n\pi_n\to\infty$.

\begin{lemma}\label{lem:clt}
Assume \textup{(A1)} and \textup{(A2)}, and let $(\wnt)$ be a random weighting with second moment $\taun$, with $\E w_{n1}^4/(n\taun^2)\to0$ and $1/\taun\to\lambda\in[0,1]$. Then, for every $\theta\in\R^m$,
\[
\E\big[\exp\{i\theta^\T(n/\taun)^{1/2}A_n\}\mid\mathcal D_n\big]\pto\exp\{-(1-\lambda)\,\theta^\T\Vm\theta/2\} .
\]
When, given $\mathcal D_n$, the weights are independent copies of a nonnegative variable with mean $1$ and second moment $\hat\tau_n$, for a $\mathcal D_n$-measurable $\hat\tau_n\ge1$ with $\E[w_{n1}^4\mid\mathcal D_n]/(n\hat\tau_n^2)\pto0$ and $1/\hat\tau_n\pto\lambda$, the conclusion holds with $\hat\tau_n$ substituted for $\taun$.
\end{lemma}

We prove Lemma~\ref{lem:clt} in the appendix.

\begin{theorem}\label{thm:limit}
Let $(\wnt)$ be a random weighting with second moment $\taun$, and assume \textup{(A1)}--\textup{(A3)}. Then the following hold.
\begin{enumerate}[label=\textup{(\roman*)},leftmargin=2.4em]
\item $\Gn\pto\Qm$, and $\Prob(E_n)\to1$ for $E_n=\{\Gn\ \text{nonsingular}\}$.
\item $(n/\taun)^{1/2}A_n\dto N(0,\Vm)$.
\item $(n/\taun)^{1/2}(\bhat-\beta)\dto N(0,\Wm)$ if and only if
\begin{equation}\label{eq:D}
(n/\taun)^{1/2}\,\sbar\pto0 .
\end{equation}
\end{enumerate}
\end{theorem}

\begin{proof}
We write $T_n=(n/\taun)^{1/2}A_n$ and $U_n=(n/\taun)^{1/2}\sbar$, and $\taun/n\to0$ by (A3).

(i) By (A2), $\hat\Qm_n\pto\Qm$, and it remains to prove $\Delta_n\pto0$, where
\[
\Delta_n=\Gn-\hat\Qm_n=\frac1n\sum_{t=1}^n(\wnt-1)\zt\zt^\T .
\]
For $c_1>0$, $\Delta_n=\Delta_{n,1}+\Delta_{n,2}$ with
\[
\Delta_{n,1}=\frac1n\sum_{t=1}^n(\wnt-1)\zt\zt^\T\ind\{\|\zt\|\le c_1\},\qquad \Delta_{n,2}=\frac1n\sum_{t=1}^n(\wnt-1)\zt\zt^\T\ind\{\|\zt\|>c_1\}.
\]
Given $\mathcal D_n$, $\Delta_{n,1}$ is a sum of independent centered terms, and $\|\zt\zt^\T\|=\|\zt\|^2$, so
\[
\E\big[\|\Delta_{n,1}\|^2\mid\mathcal D_n\big]=\frac{\taun-1}{n^2}\sum_{t=1}^n\|\zt\|^4\ind\{\|\zt\|\le c_1\}\le\frac{(\taun-1)c_1^4}{n}\to0 .
\]
For $\Delta_{n,2}$,
\[
\|\Delta_{n,2}\|\le\frac1n\sum_{t=1}^n(\wnt+1)\|\zt\|^2\ind\{\|\zt\|>c_1\},
\]
and the right side has expectation $2\E[\|z_1\|^2\ind\{\|z_1\|>c_1\}]$ by stationarity and $\E\wnt=1$. Consequently, for every $\delta>0$,
\[
\limsup_{n\to\infty}\Prob(\|\Delta_n\|>\delta)\le\frac4\delta\,\E\big[\|z_1\|^2\ind\{\|z_1\|>c_1\}\big],
\]
and $\E[\|z_1\|^2\ind\{\|z_1\|>c_1\}]\to0$ as $c_1\to\infty$. Thus $\Gn\pto\Qm$, and $\Prob(E_n)\to1$ because $\Qm$ is nonsingular.

(ii) For $\phi_n(\theta)=\E[\exp(i\theta^\T T_n)\mid\mathcal D_n]$ and $\phi(\theta)=\exp(-\theta^\T\Vm\theta/2)$, $\phi_n(\theta)\pto\phi(\theta)$ for every $\theta\in\R^m$ by Lemma~\ref{lem:clt} with $\lambda=0$. As $|\phi_n(\theta)|\le1$, $\E\exp(i\theta^\T T_n)=\E\phi_n(\theta)\to\phi(\theta)$ by bounded convergence, so $T_n\dto N(0,\Vm)$.

(iii) Proposition~\ref{prop:identity}(i) expresses $\Gn(n/\taun)^{1/2}(\bhat-\beta)$ as $T_n+U_n$ on $E_n$. Since $\Prob(E_n)\to1$ by (i), we may write
\[
\Gn(n/\taun)^{1/2}(\bhat-\beta)=T_n+U_n+o_{\Prob}(1) .
\]

Under \eqref{eq:D}, $U_n\pto0$ and $T_n+U_n\dto N(0,\Vm)$ by (ii). Since $\Gn\pto\Qm$ by (i), $(n/\taun)^{1/2}(\bhat-\beta)\dto N(0,\Qm^{-1}\Vm\Qm^{-1})=N(0,\Wm)$.

Conversely, under the normal limit in (iii), $\Gn(n/\taun)^{1/2}(\bhat-\beta)\dto N(0,\Qm\Wm\Qm)=N(0,\Vm)$ by (i), and $T_n+U_n$ has the same limit. For $\theta\in\R^m$, as $U_n$ is $\mathcal D_n$-measurable, conditioning on $\mathcal D_n$ leads to
\[
\E\exp\{i\theta^\T(T_n+U_n)\}=\E\big[\exp(i\theta^\T U_n)\,\phi_n(\theta)\big]=\phi(\theta)\,\E\exp(i\theta^\T U_n)+r_n(\theta),
\]
with $|r_n(\theta)|\le\E|\phi_n(\theta)-\phi(\theta)|\to0$, since $\phi_n(\theta)\pto\phi(\theta)$ and $|\phi_n(\theta)|\le1$. As the left side has limit $\phi(\theta)>0$, $\E\exp(i\theta^\T U_n)\to1$ for every $\theta$, and therefore $U_n\pto0$, which is \eqref{eq:D}.
\end{proof}

Since $\E\|(n/\taun)^{1/2}\sbar\|^2=\tr(\Ln)/\taun$, condition \eqref{eq:D} holds whenever $\tr(\Ln)/\taun\to0$. When $(X_t,Y_t)_{t\ge1}$ is ergodic, $\sbar\to0$ in $L^2$, so $\tr(\Ln)=n\E\|\sbar\|^2=o(n)$. Then $\taun=\{n\max(\tr(\Ln),1)\}^{1/2}$ satisfies $\taun\to\infty$, $\taun/n\to0$, and $\tr(\Ln)/\taun\to0$, so a bandwidth with \eqref{eq:D} exists, although it depends on the unknown $\Ln$.

We also consider weights which, given $\mathcal D_n$, are independent copies of a nonnegative variable with mean $1$ and a $\mathcal D_n$-measurable second moment $\hat\tau_n\ge1$. Kernel weights \eqref{eq:kernel} with a bandwidth $\hat h_n$ computed from the data have such a conditional distribution, with $\hat\tau_n=\E K^2(\xi_1/h)/\{\E K(\xi_1/h)\}^2$ at $h=\hat h_n$, because the auxiliary sample is independent of the data.

\begin{corollary}\label{cor:random}
Assume \textup{(A1)} and \textup{(A2)}, and let $\hat\tau_n\pto\infty$ and $\E[w_{n1}^4\mid\mathcal D_n]/(n\hat\tau_n^2)\pto0$. Then (i), (ii), and (iii) of Theorem~\ref{thm:limit} hold with $\hat\tau_n$ substituted for $\taun$, and \eqref{eq:D} holds for $\hat\tau_n$ whenever it holds for a deterministic sequence $\tau_n^-$ with $\Prob(\hat\tau_n\ge\tau_n^-)\to1$.
\end{corollary}

Corollary~\ref{cor:random} is proved in the appendix.

With the residuals $\ehat=Y_t-\zt^\T\bhat$ and $F_n=\{\Gn\text{ and }\sum_{t=1}^n\wnt^2\zt\zt^\T\ehat^2\ \text{nonsingular}\}$, set
\begin{equation}\label{eq:sandwich}
\Chat=\begin{cases}\Gn^{-1}\Big(\dfrac1{n^2}\displaystyle\sum_{t=1}^n\wnt^2\zt\zt^\T\ehat^2\Big)\Gn^{-1} & \text{on }F_n,\\[6pt] I_m & \text{on }F_n^c,\end{cases}
\end{equation}
a nonsingular matrix. On $F_n$, $\Chat$ is the sandwich of \citet{White1980} computed with the realized weights, invariant under multiplication of all weights by a positive constant. Its consistency, with residuals instead of the errors, holds under (A3) and the following moment condition.
\begin{description}[leftmargin=2.6em,labelindent=0pt]
\item[(A4)] $\E\|z_1\|^4<\infty$ and $\Vm$ is positive definite.
\end{description}

On $F_n$, $(n/\taun)\Chat=\Gn^{-1}\hat\Vm_n\Gn^{-1}$ with
\[
\hat\Vm_n=\frac1n\sum_{t=1}^n\frac{\wnt^2}{\taun}\,\zt\zt^\T\ehat^2 .
\]

\begin{lemma}\label{lem:sandwich}
Assume \textup{(A1)}, \textup{(A2)}, \textup{(A4)}, and $\sbar\pto0$, and let $(\wnt)$ be a random weighting with second moment $\taun$ and $\E w_{n1}^4/(n\taun^2)\to0$. Then $\hat\Vm_n\pto\Vm$.
\end{lemma}

We prove Lemma~\ref{lem:sandwich} in the appendix.

\begin{theorem}\label{thm:feasible}
Let $(\wnt)$ be a random weighting with second moment $\taun$, assume \textup{(A1)}--\textup{(A4)} and \eqref{eq:D}, and let $R$ be a $q\times m$ matrix of rank $q$. Then $(n/\taun)\Chat\pto\Wm$,
\[
(\bhat-\beta)^\T\Chat^{-1}(\bhat-\beta)\dto\chi^2_m,\qquad
(R\bhat-R\beta)^\T(R\Chat R^\T)^{-1}(R\bhat-R\beta)\dto\chi^2_q,
\]
and $(\hat\beta^{w}_{n,j}-\beta_j)/(\Chat)_{jj}^{1/2}\dto N(0,1)$ for $j=1,\dots,m$.
\end{theorem}

\begin{proof}
With $n/\taun\to\infty$, \eqref{eq:D} implies $\sbar\pto0$, and $\hat\Vm_n\pto\Vm$ by Lemma~\ref{lem:sandwich}. By Theorem~\ref{thm:limit}(i), $\Gn^{-1}\pto\Qm^{-1}$ and $\Prob(E_n)\to1$, and $\Prob(F_n)\to1$ because $\Vm$ is nonsingular, so
\[
(n/\taun)\Chat=\Gn^{-1}\hat\Vm_n\Gn^{-1}\pto\Qm^{-1}\Vm\Qm^{-1}=\Wm .
\]

With $Z_n=(n/\taun)^{1/2}(\bhat-\beta)$, $Z_n\dto Z\sim N(0,\Wm)$ by Theorem~\ref{thm:limit}(iii). In terms of $Z_n$, the three statistics of the theorem are
\[
Z_n^\T\{(n/\taun)\Chat\}^{-1}Z_n,\qquad (RZ_n)^\T\{(n/\taun)R\Chat R^\T\}^{-1}RZ_n,\qquad \frac{e_j^\T Z_n}{\{(n/\taun)(\Chat)_{jj}\}^{1/2}} .
\]
Since $(n/\taun)\Chat\pto\Wm$ and $R\Wm R^\T$ is positive definite, $R$ having rank $q$, they converge in distribution to
\[
Z^\T\Wm^{-1}Z\sim\chi^2_m,\qquad (RZ)^\T(R\Wm R^\T)^{-1}RZ\sim\chi^2_q,\qquad e_j^\T Z/\Wm_{jj}^{1/2}\sim N(0,1).
\]
\end{proof}

With $\chi^2_{m,1-\alpha}$ the quantile of order $1-\alpha$ of the $\chi^2_m$ distribution and $z_{1-\alpha/2}$ the standard normal quantile, the region
\begin{equation}\label{eq:region}
R_n(\alpha)=\{b\in\R^m\mid (\bhat-b)^\T\Chat^{-1}(\bhat-b)\le\chi^2_{m,1-\alpha}\}
\end{equation}
has asymptotic coverage $1-\alpha$, the test of $R\beta=r$ at level $\alpha$ rejects when $(R\bhat-r)^\T(R\Chat R^\T)^{-1}(R\bhat-r)>\chi^2_{q,1-\alpha}$, and $\hat\beta^{w}_{n,j}\pm z_{1-\alpha/2}(\Chat)_{jj}^{1/2}$ is a marginal interval. Simultaneous intervals for the $m$ coefficients are the projections of $R_n(\alpha)$ on the axes.

Probabilities in Theorems~\ref{thm:limit} and~\ref{thm:feasible} refer to the joint distribution of the data and the auxiliary sample. Given the data, the coverage of $R_n(\alpha)$ over the auxiliary sample tends to $1-\alpha$ in probability under the hypotheses of Theorem~\ref{thm:feasible}, by Lemma~\ref{lem:shift}, and at a fixed second moment of the weights, the coverage equals to first order the noncentral chi-squared probability of Theorem~\ref{thm:condcov}.

Replacing $Y_t$ by $Y_t+z_t^\T b$ shifts $\bhat$ by $b$ and leaves $\Chat$ unchanged. For fixed $b_0,h\in\R^m$ and $\beta=b_0+(\taun/n)^{1/2}h$, the test of $\beta=b_0$ therefore rejects with probability tending to $\Prob\{\chi^2_m(h^\T\Wm^{-1}h)>\chi^2_{m,1-\alpha}\}$, with $\chi^2_m(u)$ a noncentral chi-squared variable with $m$ degrees of freedom and noncentrality $u\ge0$, so the test detects deviations of order $\neff^{-1/2}$. When a coordinate of the score has memory order $d$, the least-squares error is of order $n^{d-1/2}$, smaller by the factor $(\taun/n^{2d})^{1/2}$, which tends to infinity under \eqref{eq:D}.

Writing $\beta=(\beta_0,\beta_X^\T)^\T$ with intercept $\beta_0\in\R$ and slopes $\beta_X\in\R^p$, we set
\begin{gather*}
\mu_X=\E X_1,\qquad \Sigma_X=\Var(X_1),\qquad \Vm_X=\E\big[(X_1-\mu_X)(X_1-\mu_X)^\T\varepsilon_1^2\big],\qquad \Wm_X=\Sigma_X^{-1}\Vm_X\Sigma_X^{-1},\\
\bar X_n=\frac1n\sum_{t=1}^nX_t,\qquad \bar X_{n,w}=\frac{\sum_{t=1}^n\wnt X_t}{\sum_{t=1}^n\wnt}.
\end{gather*}
By block inversion of $\Qm$, $\Wm_X$ is the slope block of $\Wm$. By the identity of \citet{FrischWaugh1933} and \citet{Lovell1963} applied to the weighted problem, the slope block $\hat\beta^{w}_{n,X}$ of $\bhat$ is the weighted least-squares estimator in the regression of $Y_t$ on $X_t-\bar X_{n,w}$.

\begin{corollary}\label{cor:slopes}
Let $(\wnt)$ be a random weighting with second moment $\taun$, assume \textup{(A1)} and \textup{(A3)}, let $(X_t,Y_t)_{t\ge1}$ be ergodic, and assume
\begin{equation}\label{eq:DX}
(n/\taun)^{1/2}\,\frac1n\sum_{t=1}^n(X_t-\bar X_n)\et\pto0 .
\end{equation}
Then $(n/\taun)^{1/2}(\hat\beta^{w}_{n,X}-\beta_X)\dto N(0,\Wm_X)$. If \textup{(A4)} holds as well, the conclusions of Theorem~\ref{thm:feasible} hold for $\beta_X$ and the slope block of $\Chat$.
\end{corollary}

Corollary~\ref{cor:slopes} is proved in the appendix.

With the least-squares residuals $\tilde\varepsilon_t=Y_t-\zt^\T\bols$, we set
\[
\tilde\Vm_n=\frac1n\sum_{t=1}^n\zt\zt^\T\tilde\varepsilon_t^2,\qquad \tilde C_n=\frac{\taun}{n}\,\hat\Qm_n^{-1}\tilde\Vm_n\hat\Qm_n^{-1}
\]
when $\hat\Qm_n$ and $\tilde\Vm_n$ are nonsingular, and $\tilde C_n=I_m$ otherwise. On $F_n$, $\Chat$ is computed from the observations with positive weight, about $\neff$ in number for subsampling weights, and $\tilde C_n$ from all $n$ observations.

\begin{corollary}\label{cor:tilde}
Under the hypotheses of Theorem~\ref{thm:feasible}, $(n/\taun)\tilde C_n\pto\Wm$, and the conclusions of Theorem~\ref{thm:feasible} hold for $\tilde C_n$ as well. Under the hypotheses of Corollary~\ref{cor:slopes} and \textup{(A4)}, its conclusions hold for $\beta_X$ and the slope block of $\tilde C_n$.
\end{corollary}

\begin{proof}
With $\wnt=1$ for all $t$, the weighted estimator is $\bols$ and $\hat\Vm_n=\tilde\Vm_n$, and $\E w_{n1}^4/n\to0$. Lemma~\ref{lem:sandwich} then yields $\tilde\Vm_n\pto\Vm$ under (A1), (A2), (A4), and $\sbar\pto0$. Under the hypotheses of Theorem~\ref{thm:feasible}, $\sbar\pto0$ by \eqref{eq:D} and $n/\taun\to\infty$, and under those of Corollary~\ref{cor:slopes}, (A2) and $\sbar\to0$ hold by ergodicity. Hence $(n/\taun)\tilde C_n=\hat\Qm_n^{-1}\tilde\Vm_n\hat\Qm_n^{-1}\pto\Wm$. The proofs of Theorem~\ref{thm:feasible} and Corollary~\ref{cor:slopes} depend on $\Chat$ through $(n/\taun)\Chat\pto\Wm$.
\end{proof}

By Proposition~\ref{prop:identity}(i), $\bhat-\bols=\Gn^{-1}A_n+(\Gn^{-1}-\hat\Qm_n^{-1})\sbar$ when $\Gn$ and $\hat\Qm_n$ are nonsingular, and the last term is $o_{\Prob}((\taun/n)^{1/2})$ under \eqref{eq:D}. Given the data, $(n/\taun)^{1/2}\Gn^{-1}A_n$ converges in distribution to $N(0,\Wm)$, by Theorem~\ref{thm:limit}(i) and (ii). To first order, $\bhat$ is thus least squares perturbed by an independent noise with covariance $(\taun/n)\Wm$, and $\tilde C_n$ estimates the covariance without the weights. With $\hat\Wm_n=\hat\Qm_n^{-1}\tilde\Vm_n\hat\Qm_n^{-1}$ and $\zeta\sim N(0,I_m)$ independent of the data, the region of the same shape centered at $\bols+\{(\taun-1)/n\}^{1/2}\hat\Wm_n^{1/2}\zeta$ has the asymptotic coverage and the width of $R_n(\alpha)$ with $\tilde C_n$. The factor $\taun-1$ is the factor of the conditional covariance of $A_n$ in Proposition~\ref{prop:identity}(ii). The weights draw the noise from the scores, and the first-order theory depends on them through $\taun$.

A strictly stationary sequence $(u_t)$ with mean $0$, finite variance, and autocovariances $\gamma_u(k)=\Cov(u_1,u_{1+k})$ has short memory, or memory of order $0$, if $\sum_{k\ge0}|\gamma_u(k)|<\infty$, and long memory of order $d\in(0,1/2)$ with constant $c>0$ if $\gamma_u(k)\sim c\,k^{2d-1}$ as $k\to\infty$, a special case of the time-domain definition of \citet{PipirasTaqqu2017}.

\begin{proposition}\label{prop:rates}
Assume \textup{(A1)}, let $v\in\R^m$, and let $\taun\ge1$ with $\taun\to\infty$. Then the following hold.
\begin{enumerate}[label=\textup{(\roman*)},leftmargin=2.4em]
\item If $(v^\T\st)$ has short memory, then $(n/\taun)^{1/2}v^\T\sbar\pto0$.
\item If $(v^\T\st)$ has long memory of order $d$ with constant $c$, then $\Var(\sqrt n\,v^\T\sbar)\sim c\,n^{2d}/\{d(2d+1)\}$, and $(n/\taun)^{1/2}v^\T\sbar\pto0$ whenever $\taun/n^{2d}\to\infty$.
\item Under the hypothesis of \textup{(ii)}, if $n^{1/2-d}v^\T\sbar$ converges in distribution to $Y$ with $\Prob(Y\ne0)>0$, then $(n/\taun)^{1/2}v^\T\sbar\pto0$ only if $\taun/n^{2d}\to\infty$.
\item Let $p=1$, and let $(X_t)$ and $(\et)$ be independent sequences, with $(X_t-\mu_X)$ of memory order $d_X$ and $(\et)$ of memory order $d_\varepsilon$. If $d_X+d_\varepsilon<1/2$, then $((X_t-\mu_X)\et)$ has short memory, and \eqref{eq:DX} holds. If $d_X+d_\varepsilon>1/2$, then $((X_t-\mu_X)\et)$ has long memory of order $d_X+d_\varepsilon-1/2$, and \eqref{eq:DX} holds whenever $\taun/n^{2(d_X+d_\varepsilon)-1}\to\infty$. If $d_X+d_\varepsilon=1/2$, then \eqref{eq:DX} holds whenever $\taun/\log n\to\infty$.
\end{enumerate}
\end{proposition}

We prove Proposition~\ref{prop:rates} in the appendix. For $v=e_1$, the projection is $v^\T\st=\et$, and the hypothesis of (iii) holds with a normal limit when $(\et)$ is a Gaussian or linear sequence with long memory (see \citet{Davydov1970} and \citet{Taqqu1975}). With $p=1$, $d_X+d_\varepsilon>1/2$, and $v=(-\mu_X,1)^\T$, the projection is $v^\T\st=(X_t-\mu_X)\et$, and the hypothesis holds with a non-normal limit in the Gaussian regression model of \citet{Chung2002}. Part (iv) assumes one regressor independent of the errors, and under dependence between regressors and errors, (ii) and (iii) apply to each coordinate of the score with its own memory order.

For $d_X=0$ and $d_\varepsilon>0$ in case (iv), \eqref{eq:DX} holds for every $\taun\to\infty$, and \eqref{eq:D} implies $(n/\taun)^{1/2}\bar\varepsilon_n\pto0$. When $n^{1/2-d_\varepsilon}\bar\varepsilon_n$ has a normal limit with positive variance, (iii) with $v=e_1$ then yields $\taun/n^{2d_\varepsilon}\to\infty$ and $\neff=o(n^{1-2d_\varepsilon})$.

In the models of \citet{Chung2002} and \citet{GuoKoul2008} with $d_X+d_\varepsilon>1/2$, the least-squares estimator of $\beta_X$ converges at the rate $n^{1-d_X-d_\varepsilon}$ to a non-normal limit. Under $\taun/n^{2(d_X+d_\varepsilon)-1}\to\infty$, the randomly weighted estimator $\hat\beta^{w}_{n,X}$ converges at the rate $(n/\taun)^{1/2}$ to a normal limit, by (iv) and Corollary~\ref{cor:slopes}. Their ratio is $\{\taun/n^{2(d_X+d_\varepsilon)-1}\}^{1/2}$, and $\taun$ may be chosen with the ratio diverging arbitrarily slowly, and then $\neff=o(n^{2-2d_X-2d_\varepsilon})$.

We choose $\taun$ from the data through an estimate of the memory order of the score. For $0\le a_0<1$, $\delta>0$, $a_1<1$, and a $\mathcal D_n$-measurable $\hat d_n$, let
\[
\hat a_n=\min\{\max(a_0,2\max(\hat d_n,0)+\delta),a_1\},\qquad \hat\tau_n=n^{\hat a_n},
\]
and, given $\mathcal D_n$, let the weights be the subsampling weights \eqref{eq:subsample} with $\pi_n=1/\hat\tau_n$. We compute $\tilde C_n$ with $\taun=\hat\tau_n$.

\begin{corollary}\label{cor:plugin}
With $\hat\tau_n$ and the weights as above, the following hold.
\begin{enumerate}[label=\textup{(\roman*)},leftmargin=2.4em]
\item Assume \textup{(A1)}, \textup{(A2)}, and \textup{(A4)}. Let each coordinate of $(\st)$ have short memory or long memory, let $d\in[0,1/2)$ be the largest of their orders, and let $\hat d_n\pto d$ and $2d+\delta<a_1$. Then the conclusions of Theorem~\ref{thm:feasible} hold for $\tilde C_n$ with $\taun=\hat\tau_n$.
\item Let $(X_t,Y_t)_{t\ge1}$ be ergodic and assume \textup{(A1)} and \textup{(A4)}. Let each coordinate of $((X_t-\mu_X)\et)$ and of $(X_t-\mu_X)$, and $(\et)$, have short memory or long memory, with largest orders $d$, $d_X$, and $d_\varepsilon$, and let $\hat d_n\pto d^\ast=\max(d,d_X+d_\varepsilon-1/2)$ and $2d^\ast+\delta<a_1$. Then the conclusions of Corollary~\ref{cor:slopes} hold with $\taun=\hat\tau_n$, for $\beta_X$ and the slope block of $\tilde C_n$.
\end{enumerate}
\end{corollary}

\begin{proof}
With $\delta<a_1$,
\[
\delta\le\hat a_n\le a_1<1\qquad\text{and}\qquad n^{\delta}\le\hat\tau_n\le n^{a_1},
\]
so $\hat\tau_n\to\infty$ and $\hat\tau_n/n\to0$. For the subsampling weights, $\E[w_{n1}^4\mid\mathcal D_n]=\hat\tau_n^3$, and
\[
\frac{\E[w_{n1}^4\mid\mathcal D_n]}{n\hat\tau_n^2}=\frac{\hat\tau_n}{n}\to0 .
\]
The hypotheses of Corollary~\ref{cor:random} on the weights then hold.

(i) Set $\tau_n^-=n^{2d+\delta/2}$. With $2d+\delta/2<a_1$, $\hat\tau_n\ge\tau_n^-$ whenever $\hat d_n\ge d-\delta/4$, and $\Prob(\hat d_n\ge d-\delta/4)\to1$ by the consistency of $\hat d_n$, so $\Prob(\hat\tau_n\ge\tau_n^-)\to1$. By Proposition~\ref{prop:rates}(i) and (ii) applied to each coordinate of the score, \eqref{eq:D} holds for $\tau_n^-$, since $\tau_n^-/n^{2d}\to\infty$. Corollary~\ref{cor:random} then yields \eqref{eq:D} for $\hat\tau_n$ and
\[
(n/\hat\tau_n)^{1/2}(\bhat-\beta)\dto N(0,\Wm).
\]
For the covariance, \eqref{eq:D} for $\tau_n^-$ and $n/\tau_n^-\to\infty$ imply $\sbar\pto0$, and the argument in the proof of Corollary~\ref{cor:tilde} yields
\[
(n/\hat\tau_n)\tilde C_n=\hat\Qm_n^{-1}\tilde\Vm_n\hat\Qm_n^{-1}\pto\Wm .
\]
The statistics of Theorem~\ref{thm:feasible} then converge by the argument in the proof of the theorem.

(ii) Set $\tau_n^-=n^{2d^\ast+\delta/2}$. By the argument of (i), $\Prob(\hat\tau_n\ge\tau_n^-)\to1$. The proof of Corollary~\ref{cor:slopes} depends on the weights through Theorem~\ref{thm:limit}(i) and (ii), valid with $\hat\tau_n$ by Corollary~\ref{cor:random}, and through the bound $\bar X_{n,w}-\bar X_n=O_{\Prob}((\hat\tau_n/n)^{1/2})$, obtained by the computation of the same proof. It remains to prove \eqref{eq:DX} with $\hat\tau_n$.

Whenever $\hat\tau_n\ge\tau_n^-$, the norm of the left side of \eqref{eq:DX} is at most
\[
\Big(\frac{n}{\tau_n^-}\Big)^{1/2}\Big\|\frac1n\sum_{t=1}^n(X_t-\mu_X)\et\Big\|+\Big(\frac{n}{\tau_n^-}\Big)^{1/2}\|\bar X_n-\mu_X\|\,|\bar\varepsilon_n| .
\]
The first term tends to $0$ in probability by Proposition~\ref{prop:rates}(i) and (ii), applied to each coordinate of $(X_t-\mu_X)\et$. For the second term, the computations in the proof of Proposition~\ref{prop:rates}(i) and (ii) yield
\[
\Var(\sqrt n\,\bar X_{n,k})=O(n^{2d_X})\qquad\text{and}\qquad \Var(\sqrt n\,\bar\varepsilon_n)=O(n^{2d_\varepsilon})
\]
for each coordinate $\bar X_{n,k}$ of $\bar X_n$. The second term is then $O_{\Prob}(n^{d_X+d_\varepsilon-1/2-d^\ast-\delta/4})$, with a negative exponent, since $d^\ast\ge d_X+d_\varepsilon-1/2$. Hence \eqref{eq:DX} holds with $\hat\tau_n$.

Under ergodicity, $\sbar\to0$ almost surely, so $(n/\hat\tau_n)\tilde C_n\pto\Wm$ by the argument of (i), and the conclusions of Corollary~\ref{cor:slopes} follow by the argument in the proof of the corollary.
\end{proof}

With the margin $\delta$, consistency of $\hat d_n$ implies the rate condition, at any rate of convergence of $\hat d_n$. For observed series, the local Whittle estimates of \citet{Robinson1995} are consistent for linear processes under the conditions of the same paper and, under the conditions of \citet{DallaGiraitisHidalgo2006}, for a class of nonlinear processes including nonlinear transforms of a Gaussian process. In practice, we compute the estimates from $\zt\tilde\varepsilon_t$ in (i), and from $(X_t-\bar X_n)\tilde\varepsilon_t$, $X_t-\bar X_n$, and $\tilde\varepsilon_t$ in (ii), and we assume the estimates from the residual series to be consistent. With $a_0=1/2$, $\delta=0.1$, and $a_1=0.9$, the values of the simulations and the application, the condition $2d+\delta<a_1$ is $d<0.4$, and the local Whittle estimates have $\lfloor n^{0.6}\rfloor$ frequencies.

\section{Coverage and width at a fixed second moment of the weights}\label{sec:cost}

When $\taun=\tau$, the two terms $(\tau-1)\Vm$ and $\Ln$ of the variance identity in Proposition~\ref{prop:identity}(iii) are of the same order, and so are $\sqrt n\,A_n$ and $\sqrt n\,\sbar$ in the decomposition of Proposition~\ref{prop:identity}(i). We assume the following, in the role of (A3) and \eqref{eq:D}.
\begin{description}[leftmargin=2.6em,labelindent=0pt]
\item[(A5)] $\wnt=w_t$, with $w_1,w_2,\dots$ independent and identically distributed, $\E w_1=1$, $\E w_1^2=\tau>1$, and $\E w_1^4<\infty$.
\item[(A6)] $\sqrt n\,\sbar=O_{\Prob}(1)$.
\end{description}
Condition (A6) holds when $\tr(\Ln)$ is bounded, in particular when the autocovariances of the score are summable.

For a random weighting with second moment $\taun$, we set $\Omega_n=(n/\taun)^{1/2}(\bols-\beta)$ and, for $\lambda\in[0,1)$ and $x>0$,
\[
\Pi_x(b)=\Prob\Big\{\chi^2_m\Big(\frac{b^\T\Wm^{-1}b}{1-\lambda}\Big)\le\frac{x}{1-\lambda}\Big\},\qquad b\in\R^m .
\]

\begin{lemma}\label{lem:shift}
Let $(\wnt)$ be a random weighting with second moment $\taun$, with $\E w_{n1}^4/(n\taun^2)\to0$ and $1/\taun\to\lambda\in[0,1)$, and assume \textup{(A1)}, \textup{(A2)}, \textup{(A4)}, and $(n/\taun)^{1/2}\sbar=O_{\Prob}(1)$. Then, for every $x>0$,
\[
\Prob\{(\bhat-\beta)^\T\Chat^{-1}(\bhat-\beta)\le x\mid\mathcal D_n\}-\Pi_x(\Omega_n)\pto0 .
\]
\end{lemma}

Lemma~\ref{lem:shift} is proved in the appendix. Under (A3) and \eqref{eq:D}, the hypotheses of the lemma hold with $\lambda=0$ and $\Omega_n\pto0$, and $\Pi_x$ is continuous, so $\Prob\{(\bhat-\beta)^\T\Chat^{-1}(\bhat-\beta)\le x\mid\mathcal D_n\}\pto\Pi_x(0)=\Prob(\chi^2_m\le x)$. For $\taun=\tau$, the hypotheses hold with $\lambda=1/\tau$ by (A5) and (A6), and $\Pi_x(b)=\Psi_x\{\tau b^\T\Wm^{-1}b/(\tau-1)\}$ with
\[
\Psi_x(u)=\Prob\Big\{\chi^2_m(u)\le\frac{\tau x}{\tau-1}\Big\},\qquad u\ge0 .
\]
For fixed $x$, $\Psi_x$ decreases strictly and continuously from $\Psi_x(0)$ to $\lim_{u\to\infty}\Psi_x(u)=0$, and we set
\[
\Lambda_n=\frac{n\,\sbar^\T\Vm^{-1}\sbar}{\tau-1},\qquad \E\Lambda_n=\frac{\tr(\Vm^{-1}\Ln)}{\tau-1}.
\]

\begin{theorem}\label{thm:condcov}
Assume \textup{(A1)}, \textup{(A2)}, and \textup{(A4)}--\textup{(A6)}. Then the following hold.
\begin{enumerate}[label=\textup{(\roman*)},leftmargin=2.4em]
\item $n\Chat\pto\tau\Wm$.
\item For every $x>0$,
\[
\Prob\{(\bhat-\beta)^\T\Chat^{-1}(\bhat-\beta)\le x\mid\mathcal D_n\}-\Psi_x(\Lambda_n)\pto0 .
\]
\end{enumerate}
\end{theorem}

\begin{proof}
(i) Since $\tau/n\to0$, the argument of Theorem~\ref{thm:limit}(i) yields $\Gn\pto\Qm$ and $\Prob(E_n)\to1$. Under (A6), $\sbar\pto0$, and $\E w_1^4<\infty$, so Lemma~\ref{lem:sandwich} applies and $\hat\Vm_n\pto\Vm$. On $F_n$, $n\Chat=\Gn^{-1}\{\tau\hat\Vm_n\}\Gn^{-1}$, and $\Prob(F_n)\to1$, whence $n\Chat\pto\tau\Qm^{-1}\Vm\Qm^{-1}=\tau\Wm$.

(ii) By Lemma~\ref{lem:shift} with $\lambda=1/\tau$, $\Prob\{(\bhat-\beta)^\T\Chat^{-1}(\bhat-\beta)\le x\mid\mathcal D_n\}=\Psi_x\{\tau \Omega_n^\T\Wm^{-1}\Omega_n/(\tau-1)\}+o_{\Prob}(1)$. Since $\bols-\beta=\hat\Qm_n^{-1}\sbar$ whenever $\hat\Qm_n$ is nonsingular, $\hat\Qm_n\pto\Qm$, $\Qm\Wm^{-1}\Qm=\Vm^{-1}$, and $\sqrt n\,\sbar=O_{\Prob}(1)$, we have
\[
\frac{\tau \Omega_n^\T\Wm^{-1}\Omega_n}{\tau-1}=\frac{n(\bols-\beta)^\T\Wm^{-1}(\bols-\beta)}{\tau-1}=\Lambda_n+o_{\Prob}(1),
\]
and $\Psi_x$ is uniformly continuous on $[0,\infty)$, so $\Psi_x\{\tau \Omega_n^\T\Wm^{-1}\Omega_n/(\tau-1)\}-\Psi_x(\Lambda_n)\pto0$.
\end{proof}

When $\Lambda_n$ has a limit distribution, the coverage converges, and we assume the following central limit theorem for the score mean, which implies (A6).
\begin{description}[leftmargin=2.6em,labelindent=0pt]
\item[(A7)] $\sqrt n\,\sbar\dto N(0,\Lm)$ for a covariance matrix $\Lm$.
\end{description}
Since $\Vm$ is positive definite by (A4), $\Vm^{-1}\Lm$ has the eigenvalues $\kappa_1\ge\dots\ge\kappa_m\ge0$ of the symmetric nonnegative definite matrix $\Vm^{-1/2}\Lm\Vm^{-1/2}$. For $v\ne0$, set
\[
\kappa(v)=\frac{v^\T\Lm v}{v^\T\Vm v}\in[\kappa_m,\kappa_1].
\]
For serially uncorrelated scores, $\Lm=\Vm$ and $\kappa_1=\dots=\kappa_m=1$. When the autocovariances of the score are summable, $\kappa(v)>1$ if and only if the autocovariances of $(v^\T\st)$ at positive lags have a positive sum. By (A7), \[\Lambda_n\dto\sum_{j=1}^m\kappa_j\zeta_j^2/(\tau-1),\] where $\zeta_1,\dots,\zeta_m$ are independent standard normal variables. At fixed $\tau$, the limit distribution of $\sqrt n(\bhat-\beta)$ depends on $\Lm$ through the variance identity of Proposition~\ref{prop:identity}(iii), and $n\Chat$ estimates $\tau\Wm$ by Theorem~\ref{thm:condcov}(i), so the limit distribution of the Wald statistic depends on $\kappa_1,\dots,\kappa_m$.

\begin{proposition}\label{prop:fixedtau}
Assume \textup{(A1)}, \textup{(A2)}, \textup{(A4)}, \textup{(A5)}, and \textup{(A7)}. Then the following hold.
\begin{enumerate}[label=\textup{(\roman*)},leftmargin=2.4em]
\item $\sqrt n(\bhat-\beta)\dto N\big(0,\Qm^{-1}\{(\tau-1)\Vm+\Lm\}\Qm^{-1}\big)$.
\item $(\bhat-\beta)^\T\Chat^{-1}(\bhat-\beta)\dto\sum_{j=1}^ma_j\zeta_j^2$, where $a_j=(\tau-1+\kappa_j)/\tau$.
\item For $c\in\R^m$, $c\ne0$, and $v=\Qm^{-1}c$, $c^\T(\bhat-\beta)/(c^\T\Chat c)^{1/2}\dto N\big(0,(\tau-1+\kappa(v))/\tau\big)$.
\end{enumerate}
\end{proposition}

We prove Proposition~\ref{prop:fixedtau} in the appendix and derive from it the coverage of $R_n(\alpha)$ and the width of the marginal intervals relative to least squares. Under a bound on the ratio of the long-run to the marginal variance of the score, we also derive a choice of $\tau$. We set $a_{\max}=(\tau-1+\kappa_1)/\tau$, $a_{\min}=(\tau-1+\kappa_m)/\tau$, and $v_j=\Qm^{-1}e_j$ for $j=1,\dots,m$.

\begin{corollary}\label{cor:cost}
Assume the hypotheses of Proposition~\ref{prop:fixedtau}. Then the following hold.
\begin{enumerate}[label=\textup{(\roman*)},leftmargin=2.4em]
\item The limit of $\Prob\{\beta\in R_n(\alpha)\}$ exists and satisfies
\[
\Prob\Big(\chi^2_m\le\frac{\chi^2_{m,1-\alpha}}{a_{\max}}\Big)\le\lim_n\Prob\{\beta\in R_n(\alpha)\}\le\Prob\Big(\chi^2_m\le\frac{\chi^2_{m,1-\alpha}}{a_{\min}}\Big).
\]
For each $j$, the marginal interval $\hat\beta^{w}_{n,j}\pm z_{1-\alpha/2}(\Chat)_{jj}^{1/2}$ has limit coverage $\Prob[|\zeta_1|\le z_{1-\alpha/2}\{\tau/(\tau-1+\kappa(v_j))\}^{1/2}]$.
\item $\sqrt n\,e_j^\T(\bols-\beta)\dto N(0,\kappa(v_j)\,v_j^\T\Vm v_j)$.
\item Let $\kbar>1$ and $\rhostar>0$. If $\kappa_1\le\kbar$ and $\tau\ge(\kbar-1)/\rhostar$, then $a_{\max}\le1+\rhostar$, the region $\{b\in\R^m\mid (\bhat-b)^\T\Chat^{-1}(\bhat-b)\le(1+\rhostar)\chi^2_{m,1-\alpha}\}$ has limit coverage at least $1-\alpha$, and $\neff\le\rhostar n/(\kbar-1)$.
\end{enumerate}
\end{corollary}

Corollary~\ref{cor:cost} is proved in the appendix. By Proposition~\ref{prop:fixedtau}(i) and Corollary~\ref{cor:cost}(ii), the asymptotic standard deviations of $\hat\beta^{w}_{n,j}$ and of $e_j^\T\bols$ are in the ratio $\{(\tau-1+\kappa(v_j))/\kappa(v_j)\}^{1/2}$ when $\kappa(v_j)>0$. In (iii), $a_{\max}-1$ is the largest relative understatement of the variance by the sandwich over the directions of $\R^m$. The bound $\kbar$ concerns the long-run covariance of the score, not consistently estimable without restrictions on the dependence (see \citet{KojevnikovSong2023}), so (iii) applies when $\kbar$ comes from a model or from other data.

Among constructions less dependent on the auxiliary sample, centering $R_n(\alpha)$ at least squares yields
\[
R_n^{\mathrm{ols}}(\alpha)=\{b\in\R^m\mid (\bols-b)^\T\Chat^{-1}(\bols-b)\le\chi^2_{m,1-\alpha}\}.
\]
Averaging the estimator over $B$ independent draws $w^{(1)},\dots,w^{(B)}$ of the weights produces $\bar\beta_{n,B}=B^{-1}\sum_{\ell=1}^B\hat\beta^{(\ell)}_n$, where $\hat\beta^{(\ell)}_n$ is the estimator for the draw $w^{(\ell)}$. With the averaged weights $\bar w_t=B^{-1}\sum_{\ell=1}^Bw^{(\ell)}_t$, a random weighting with second moment $\tau_B=1+(\taun-1)/B$, we write $\hat\beta^{\bar w}_n$ and $\hat C^{\bar w}_n$ for the estimator \eqref{eq:estimator} and the sandwich \eqref{eq:sandwich} computed from $\bar w$. Centering at the average with the sandwich $\hat C^{\bar w}_n$ leads to
\[
\bar R_{n,B}(\alpha)=\{b\in\R^m\mid (\bar\beta_{n,B}-b)^\T(\hat C^{\bar w}_n)^{-1}(\bar\beta_{n,B}-b)\le\chi^2_{m,1-\alpha}\}.
\]

From $J$ independent draws, \citet{Song2016} built a randomized confidence function, used by \citet{Leung2022}. For a slack $\nu\in(0,\alpha)$, $x_\nu=\chi^2_{m,1-\alpha+\nu}$, and the Wald statistic $S_n(b;w)=(\bhat-b)^\T\Chat^{-1}(\bhat-b)$ at $b$ computed from the draw $w$, the construction and its limit are
\[
C_{n,J}=\Big\{b\in\R^m\,\Big|\,\frac1J\sum_{\ell=1}^J\ind\{S_n(b;w^{(\ell)})\le x_\nu\}\ge1-\alpha\Big\},\qquad
C_{n,\infty}=\{b\in\R^m\mid p_n(b)\ge1-\alpha\},
\]
where $p_n(b)=\Prob\{S_n(b;w)\le x_\nu\mid\mathcal D_n\}$. Given $\mathcal D_n$, the draws are independent, so $\ind\{b\in C_{n,J}\}\to\ind\{b\in C_{n,\infty}\}$ almost surely as $J\to\infty$ for every $b$ with $p_n(b)\ne1-\alpha$. Since $\Psi_{x_\nu}(0)>\Prob(\chi^2_m\le x_\nu)=1-\alpha+\nu$, the equation $\Psi_{x_\nu}(u)=1-\alpha$ has a unique solution $u=\Lambda^\ast>0$.

\begin{proposition}\label{prop:limits}
Let $B\ge1$ and $\nu\in(0,\alpha)$ be fixed, and let $x=\chi^2_{m,1-\alpha}$. Then the following hold.
\begin{enumerate}[label=\textup{(\roman*)},leftmargin=2.4em]
\item Under the hypotheses of Theorem~\ref{thm:feasible}, $\Prob\{\beta\in R_n^{\mathrm{ols}}(\alpha)\}\to1$, $\Prob(\beta\in C_{n,\infty})\to1$, $\Prob\{\beta\in\bar R_{n,B}(\alpha)\}\to1-\alpha$, and $\bar\beta_{n,B}-\hat\beta^{\bar w}_n=o_{\Prob}((\taun/n)^{1/2})$.
\item Under the hypotheses of Theorem~\ref{thm:condcov}, $(\bols-\beta)^\T\Chat^{-1}(\bols-\beta)=(\tau-1)\Lambda_n/\tau+o_{\Prob}(1)$, $p_n(\beta)=\Psi_{x_\nu}(\Lambda_n)+o_{\Prob}(1)$, $\bar\beta_{n,B}-\hat\beta^{\bar w}_n=o_{\Prob}(n^{-1/2})$, and
\[
\Prob\{\beta\in\bar R_{n,B}(\alpha)\mid\mathcal D_n\}-\Prob\Big\{\chi^2_m(B\Lambda_n)\le\frac{\tau_Bx}{\tau_B-1}\Big\}\pto0 .
\]
\item Under the hypotheses of Proposition~\ref{prop:fixedtau}, the coverages of $R_n^{\mathrm{ols}}(\alpha)$, $C_{n,\infty}$, and $\bar R_{n,B}(\alpha)$ converge to
\[
\Prob\Big(\sum_{j=1}^m\kappa_j\zeta_j^2/\tau\le x\Big),\qquad \Prob\Big\{\sum_{j=1}^m\kappa_j\zeta_j^2\le(\tau-1)\Lambda^\ast\Big\},\qquad \Prob\Big\{\sum_{j=1}^ma_j(B)\zeta_j^2\le x\Big\},
\]
where $a_j(B)=(\tau-1+B\kappa_j)/(\tau-1+B)$, the first limit being at least $\lim_n\Prob\{\beta\in R_n(\alpha)\}$, and at least $1-\alpha$ when $\kappa_1\le\tau$.
\end{enumerate}
\end{proposition}

We prove Proposition~\ref{prop:limits} in the appendix. By (i), $R_n^{\mathrm{ols}}(\alpha)$, of the shape and size of $R_n(\alpha)$, has asymptotic coverage one when $\taun\to\infty$, so the nominal level of $R_n(\alpha)$ comes from the randomization of the center, and \citet{Cox1958} and \citet{GeyerMeeden2005} describe a similar exchange for randomized confidence sets.

At fixed $\tau$, by (ii), the region averaged over $B$ draws has the conditional coverage of a single draw with second moment $\tau_B$, and $R_n^{\mathrm{ols}}(\alpha)$ and $C_{n,\infty}$ are to first order ellipsoids centered at least squares, $\{b\in\R^m\mid n(\bols-b)^\T\Wm^{-1}(\bols-b)\le\tau x\}$ and the same set with $(\tau-1)\Lambda^\ast$ in the role of $\tau x$. Under (A7), $a_j(B)\to\kappa_j$ as $B\to\infty$ in (iii), the coefficients of the limit distribution $\sum_j\kappa_j\zeta_j^2$ of the least-squares Wald statistic with the covariance of \citet{White1980}.

By Lemma~\ref{lem:sandwich} with $\wnt=1$ for every $t$, the limits $(n/\taun)\Chat\pto\Wm$ and $n\Chat\pto\tau\Wm$ hold with $\tilde C_n$ in the role of $\Chat$, and at $\taun=\tau_B$ with $(\tau_B/n)\hat\Qm_n^{-1}\tilde\Vm_n\hat\Qm_n^{-1}$ in the role of $\hat C^{\bar w}_n$. Lemma~\ref{lem:shift}, Theorem~\ref{thm:condcov}, Propositions~\ref{prop:fixedtau} and~\ref{prop:limits}, and Corollary~\ref{cor:cost} then hold with the same substitutions, the proofs depending on the covariance estimates through the limits.

\section{Simulations}\label{sec:evidence}

\subsection{Designs and methods}\label{sec:sim-structures}

Let
\[
Y_t=\beta_0+\beta_1X_{1t}+\beta_2X_{2t}+\et,\qquad \beta=(1,1,-1)^\T,
\]
with $(\et)$ independent of $(X_{1t},X_{2t})$ and
\[
X_{1t}=\begin{cases}0.5X_{1,t-1}+(3/4)^{1/2}\eta_{1t} & \text{(AR(1))},\\[2pt] \kappa(1-L)^{-d_X}\eta_{1t} & \text{(ARFIMA)},\end{cases}\qquad X_{2t}=-0.5X_{2,t-1}+(3/4)^{1/2}\eta_{2t},
\]
where $L$ is the lag operator, $\kappa=\{\Var(1-L)^{-d_X}\eta_{1t}\}^{-1/2}$, and $(\eta_{1t},\eta_{2t})$ are independent $N\bigl(0,\bigl(\begin{smallmatrix}1&0.5\\0.5&1\end{smallmatrix}\bigr)\bigr)$ vectors. Errors and parameter values are in Table~\ref{tab:designs}, with $(u_t)$ independent $N(0,1)$ variables and $(G_t)$ fractional Gaussian noise (fGn) of Hurst index $H$.

\begin{table}[!ht]
\caption{Designs and memory order $d$ of the score coordinate under study.}
\label{tab:designs}
\centering\small
\setlength{\tabcolsep}{6pt}
\begin{tabular}{l l l l l}
\toprule
design & $\et$ & $X_{1t}$ & parameter & $d$\\
\midrule
AR(1) errors & $\phi\varepsilon_{t-1}+u_t$ & AR(1) & $\phi=0.3,\ 0.6,\ 0.9$ & $0$\\
squared fGn errors & $(G_t^2-1)/\sqrt2$ & AR(1) & $H=0.7,\ 0.8,\ 0.85$ & $0,\ 0.1,\ 0.2$\\
ARFIMA errors & $(1-L)^{-d_\varepsilon}u_t$ & AR(1) & $d_\varepsilon=0.1,\ 0.2,\ 0.3$ & $0.1,\ 0.2,\ 0.3$\\
joint long memory & $(1-L)^{-d_\varepsilon}u_t$ & ARFIMA & $d_X=d_\varepsilon=0.2,\ 0.3,\ 0.35$ & $0,\ 0.1,\ 0.2$\\
\bottomrule
\end{tabular}
\end{table}

In the designs of Table~\ref{tab:designs}, assumptions (A1), (A2), and (A4) hold, and $(X_{1t},X_{2t},\et)$ is mixing, by the result of \citet{Maruyama1949}. With an AR(1) regressor, the slope coordinates of the score have short memory whatever the errors, so the memory order $d$ of the score is the memory order of the intercept coordinate $\et$. We therefore report the interval for $\beta_0$ and the region for $\beta$, under the rate condition \eqref{eq:D}, equivalent to $\taun/n^{2d}\to\infty$ by Proposition~\ref{prop:rates}.

Under joint long memory, the intercept coordinate has memory order $d_\varepsilon$, up to $0.35$, while the slope coordinates have memory order $d=\max(d_X+d_\varepsilon-1/2,0)$, below $d_\varepsilon$. We then report the interval for $\beta_1$, under the condition for the slopes of Corollary~\ref{cor:slopes}, met for $\taun/n^{2d}\to\infty$.

We apply random weighting by subsampling, with $\tau=n^a$ for $a\in\{0.1,0.2,\dots,0.8\}$ and with the data-driven $\hat\tau_n=n^{\hat a_n}$ of Corollary~\ref{cor:plugin}, part (i) for $\beta_0$ and part (ii) for $\beta_1$. At fixed $a$, the rate condition holds if and only if $a>2d$, true at the default $a=1/2$ in every design except $d_\varepsilon=0.3$. For $\beta_j$, the coefficient under study, and for $\beta$, we form the interval and the region at level $1-\alpha$,
\[
\hat\beta^{w}_{n,j}\pm z_{1-\alpha/2}(\tilde C_n)_{jj}^{1/2}\qquad\text{and}\qquad R_n(\alpha)\ \text{with }\tilde C_n,
\]
and the same two with the sandwich $\Chat$.

For the same coefficient, we also compute, at the tuning values recommended by the authors, the resampled interval of \citet[equation~(3)]{Leung2022}, the closest to ours in construction, self-normalization (SN), and the equal-weighted cosine (EWC) interval, the fixed-bandwidth recommendation of \citet{LazarusLewisStockWatson2018}. In the resampled construction, the interval is centered at the average of $N_n=n^{1/2}$ draws with replacement from the $j$th coordinates of the pseudo-observations $\xi_t=n(Z^\T Z)^{-1}z_tY_t$, with $Z$ the regressor matrix, and has half-width $z_{1-\alpha/2}\hat\sigma_nN_n^{-1/2}$, where $\hat\sigma_n$ is the standard deviation of the coordinates. At $a=1/2$, $N_n=\neff$. With least squares converging at the rate $n^{1/2-d}$, the interval is valid for $d<1/4$ (see \citet[Remark~4]{Leung2022}). For $\beta$, the region of \citet[Section~2.1]{Leung2022} is the set of $b$ with
\[
N_n(\bar\xi^*_n-b)^\T\hat\Sigma_n^{-1}(\bar\xi^*_n-b)\le\chi^2_{m,1-\alpha},
\]
where $\bar\xi^*_n$ is the average of $N_n$ draws with replacement from the $\xi_t$ and $\hat\Sigma_n$ the covariance matrix of the $\xi_t$, and is valid for $d_{\max}<1/4$, with $d_{\max}$ the largest memory order among the score coordinates. Through the $\xi_t$, the size of the resampled interval and region depends on $\beta$, and random weighting, by equivariance, has a size independent of $\beta$.

For SN, we take the interval of \citet{Shao2010}, normalized by recursive estimates of the coefficients, and for EWC, $\lfloor0.4n^{2/3}\rfloor$ cosine terms and Student critical values with the same number of degrees of freedom. Under long memory, the limit distributions behind the critical values of SN and EWC, derived for short memory, here $d=0$, change form, a result of \citet{McElroyPolitis2012} for the mean.

For $n\in\{1000,\ 4000,\ 16{,}000,\ 64{,}000\}$, we draw $1000$ samples from the stationary distribution of the design and apply the methods to the same samples, generating long-memory series with the simulation algorithm of \citet{ChanWood1999}. For Figures~\ref{fig:compare} and S2, we also draw $1000$ samples at $n=2000$, $8000$ and $32{,}000$ and apply the methods at the recommended tuning values and random weighting at $a=1/2$. For Figure~\ref{fig:beta}, we take $\beta_1\in\{0,1,\dots,8\}$ in the samples of the design with $d_\varepsilon=0.1$ at $n=16{,}000$. An R script reproducing the simulations and the application is supplementary material.

\subsection{Results}\label{sec:sim-results}

\begin{figure}[!b]
\centering
\includegraphics[width=\textwidth]{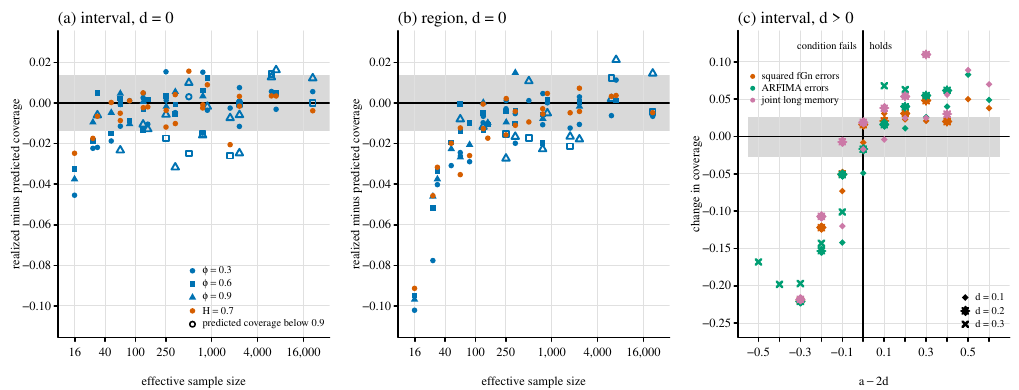}
\caption{Realized minus predicted coverage of random weighting against $\neff$ for $d=0$ (a, b), and change in coverage from $n=1000$ to $64{,}000$ for $d>0$ (c), with grey bands of two Monte Carlo standard errors.}
\label{fig:accuracy}
\end{figure}

\begin{figure}[!t]
\centering
\includegraphics[width=\textwidth]{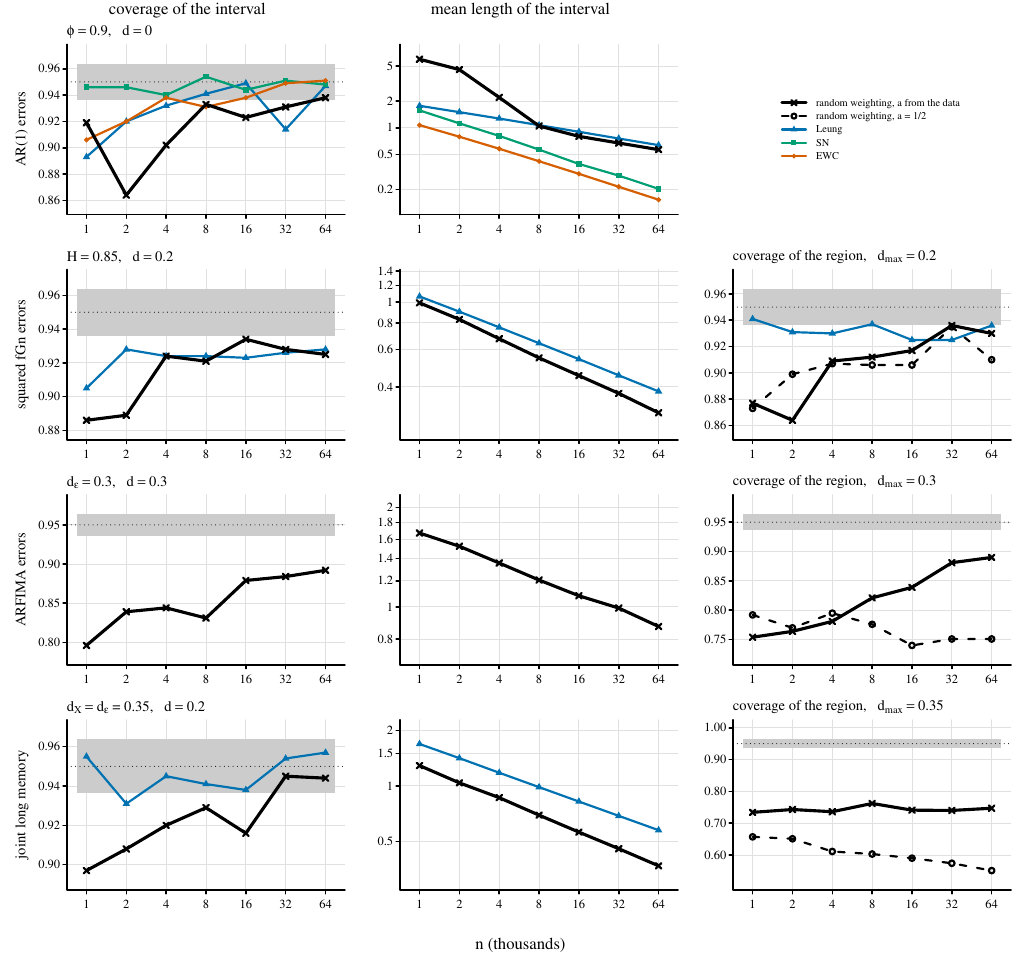}
\caption{Coverage and mean length of the $95\%$ intervals, and coverage of the $95\%$ regions, against $n$ in the design with the strongest dependence in every family.}
\label{fig:compare}
\end{figure}

For $d=0$, with (A7) holding by the central limit theorem of \citet{Arcones1994}, realized coverage in Figure~\ref{fig:accuracy}(a) and (b) follows the limits of Corollary~\ref{cor:cost}(i) and Proposition~\ref{prop:fixedtau}(ii), $0.005$ below on average for the interval and within $0.035$ for the region once $\neff$ exceeds $40$. With $\neff$ down to $4$, at $a=0.7$ and $0.8$, the interval falls short by up to $0.123$. For $\neff<50$, the sandwich $\Chat$, built from about $\neff$ observations, leaves a shortfall about three times that of $\tilde C_n$, $0.049$ against $0.017$ for the interval and $0.172$ against $0.053$ for the region. In the designs of Figure~\ref{fig:accuracy}(a) and (b), $\kappa_1\le20$, so Corollary~\ref{cor:cost}(iii) applies with $\kbar=20$ and $\rhostar=1$, at $\tau=19$ and with the critical value $2\chi^2_{m,1-\alpha}$. The region's coverage is then at least $0.979$, against $0.854$ with the usual critical value at $\phi=0.9$ and $n=1000$.

Under long memory, coverage in Figure~\ref{fig:accuracy}(c) moves toward $0.95$ when $a>2d$, in line with Theorem~\ref{thm:limit}(iii) and Corollary~\ref{cor:tilde}, and away from $0.95$ otherwise. Convergence is slow, the score mean being of order $n^{-(a-2d)/2}$ relative to the weights' noise, and coverage at $n=64{,}000$ ranges from $0.844$ to $0.950$. Regions behave the same way with $d_{\max}$ in the role of $d$, with a single exception among $56$ cells. Under joint long memory at $d_\varepsilon=0.35$, $a=1/2$ meets the condition for the slopes, and the condition for the region is $a>2d_{\max}=0.7$, so the interval for $\beta_1$ covers $0.941$ at $n=64{,}000$ and the region's coverage decreases with $n$ in Figure~\ref{fig:compare}.

Chosen from the data, the exponent equals the floor $1/2$ from $n=16{,}000$ when $d=0$ and is at most $0.577$ for $d=0.1$ and $0.2$, and a fixed exponent picked with the truth known improves average coverage by at most $0.011$. In a small sample, strong short memory raises the memory estimate, and at $\phi=0.9$ and $n=1000$, an average exponent of $0.88$ makes the interval $3.8$ times longer than at $a=1/2$. At $d_\varepsilon=0.3$, an exponent near $0.7$ leaves $11$ to $30$ effective observations, hence the slow increase of coverage. For the slope under joint long memory, part (ii) raises coverage to $0.897$ to $0.944$, from $0.820$ to $0.922$ with part (i), set by the intercept coordinate's memory order.

\begin{figure}[!t]
\centering
\includegraphics[width=\textwidth]{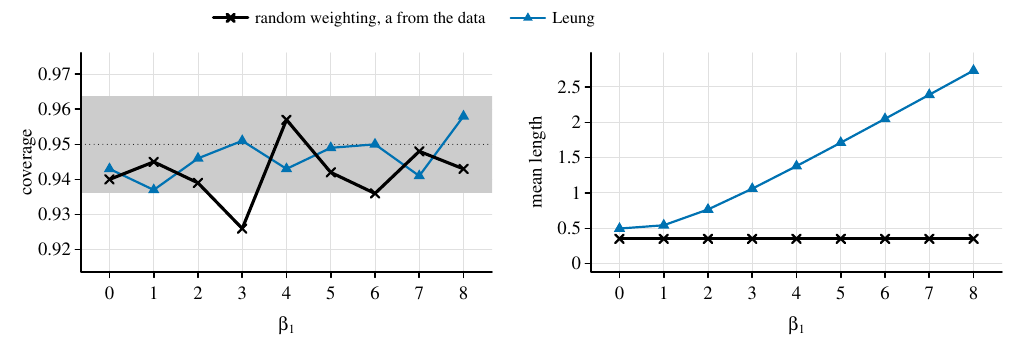}
\caption{Coverage and mean length of the $95\%$ intervals for $\beta_0$ against $\beta_1$ under ARFIMA errors with $d_\varepsilon=0.1$ at $n=16{,}000$, with a grey band of two Monte Carlo standard errors around $0.95$.}
\label{fig:beta}
\end{figure}

For $d=0.1$ and $0.2$, the interval of \citet{Leung2022}, the closest construction, has $1.1$ to $1.9$ times the length of random weighting's, in Figure~S2 of the Supplementary Material, and at equal length, the two coverages differ by at most $0.026$ once $\neff$ is about $50$ or more, in Figure~S1. With $\beta_1$ increasing to $8$ in Figure~\ref{fig:beta}, both intervals for $\beta_0$ keep coverage near $0.95$, and the resampled interval, of length dependent on the other coefficients, becomes $7.8$ times longer than random weighting, unchanged at $0.35$ by equivariance.

For the region, the exponent of part (i) exceeds $2d_{\max}$, and coverage increases with $n$ wherever $d_{\max}\le0.3$, to at least $0.929$ at $n=64{,}000$ for $d_{\max}\le0.2$ and to $0.882$ to $0.890$ for $d_{\max}=0.3$. At $d_\varepsilon=0.35$, an exponent near $0.8$ leaves $7$ to $10$ effective observations, too few for a region. Where $0<d_{\max}<1/4$, the region of \citet{Leung2022} is larger by a factor of $1.5$ to $1.8$ per coefficient, with coverage within $0.010$ of random weighting's at $n=64{,}000$.

In Table~\ref{tab:outside}, EWC's coverage decreases with $n$, because the score mean's variance comes from the lowest frequencies, near the pole of the spectral density, while EWC puts equal weights on $\nu$ cosine terms, $\nu$ of order $n^{2/3}$, and so underestimates the variance by a factor of order $\nu^{2d}$. SN's coverage depends on $d$ and changes by at most $0.031$ between $n=1000$ and $64{,}000$, the self-normalized statistic having a limit dependent on $d$. Random weighting's coverage with the data-driven exponent increases with $n$ in every design, including $d_\varepsilon=0.3$, against decreasing coverage for the interval of \citet{Leung2022} and for random weighting at $a=1/2$.

\begin{table}[!ht]
\caption{Coverage of the $95\%$ intervals for $d>0$ at $n=1000$ and $64{,}000$, by random weighting with $\hat a_n$ and by SN and EWC.}
\label{tab:outside}
\centering\small
\setlength{\tabcolsep}{5pt}
\begin{tabular}{l c cc cc cc}
\toprule
 & & \multicolumn{2}{c}{random weighting} & \multicolumn{2}{c}{SN} & \multicolumn{2}{c}{EWC}\\
\cmidrule(lr){3-4}\cmidrule(lr){5-6}\cmidrule(lr){7-8}
design & $d$ & $n=1000$ & $64{,}000$ & $n=1000$ & $64{,}000$ & $n=1000$ & $64{,}000$\\
\midrule
$H=0.8$ & $0.1$ & 0.920 & 0.947 & 0.887 & 0.901 & 0.801 & 0.729 \\
$H=0.85$ & $0.2$ & 0.886 & 0.925 & 0.821 & 0.790 & 0.664 & 0.480 \\
$d_\varepsilon=0.1$ & $0.1$ & 0.934 & 0.948 & 0.918 & 0.903 & 0.872 & 0.670 \\
$d_\varepsilon=0.2$ & $0.2$ & 0.898 & 0.911 & 0.829 & 0.831 & 0.654 & 0.426 \\
$d_\varepsilon=0.3$ & $0.3$ & 0.796 & 0.892 & 0.748 & 0.726 & 0.464 & 0.225 \\
$d_X=d_\varepsilon=0.3$ & $0.1$ & 0.932 & 0.940 & 0.957 & 0.942 & 0.915 & 0.826 \\
$d_X=d_\varepsilon=0.35$ & $0.2$ & 0.897 & 0.944 & 0.939 & 0.938 & 0.875 & 0.736 \\
\bottomrule
\end{tabular}
\end{table}

\section{Real data application to wind speeds in Ireland}\label{sec:application}

\subsection{Data and design}\label{sec:app-design}

In wind-resource assessment, the long-term mean speed at a site is the quantity of interest, usually known from a record of a few years and extended through a regression on a long record at a reference station, the measure-correlate-predict method reviewed by \citet{CartaVelazquezCabrera2013}. In the daily wind speeds of Irish synoptic stations, \citet{HaslettRaftery1989} found long memory, with standard errors too small under independence between days. We use seven stations of the original study, with near-complete hourly records over 2000--2024 in the Integrated Surface Database (see \citet{SmithLottVose2011}), and the $n=8946$ of the $9132$ days with at least eight hourly reports at every station. With $W_{st}$ the mean of the hourly speeds at station $s$ on day $t$, in metres per second, we study, following \citet{HaslettRaftery1989},
\[
V_{st}=W_{st}^{1/2}-\hat g(t),
\]
with $\hat g$ the seasonal effect. We fit $\hat g$, a function of the day of the year with three annual harmonics, to the means of $W_{st}^{1/2}$ over years and stations, and use the same $\hat g$ for every station and every block below. For station $s$, we regress $V_{st}$ on the mean of the other six stations,
\[
V_{st}=\beta_{0,s}+\beta_{1,s}x_{st}+\varepsilon_{st},\qquad x_{st}=\frac16\sum_{r\neq s}V_{rt},
\]
with parameters of interest $\mu_s=\E V_{st}$ and $\beta_{1,s}$.

We estimate the memory orders by local Whittle from the $n$ days, with $m=\lfloor n^{0.6}\rfloor=234$ frequencies and hence an asymptotic standard error of $1/(2\sqrt m)=0.033$, and report the estimates in Table~\ref{tab:memory}. The estimates for $V_{st}$ are more than two standard errors above zero at every station, and the estimates for the least-squares residuals $\tilde\varepsilon_{st}$ are larger. Regressor and errors then have long memory, the case of Corollary~\ref{cor:slopes}, with a possibly non-normal limit for least squares and an infinite long-run variance, the basis of SN and EWC. For $\beta_{1,s}$, the rate condition concerns $(x_{st}-\bar x_s)\tilde\varepsilon_{st}$, with largest estimate $0.217$, so $a>2\times0.217=0.43$ suffices, the default $a=1/2$ included.

\begin{table}[!ht]
\caption{Local Whittle estimates of the memory order from the $n$ days, with standard error $0.033$, and exponents $\hat a_n$ of Corollary~\ref{cor:plugin} for the regression on $x_{st}$.}
\label{tab:memory}
\centering\small
\setlength{\tabcolsep}{7pt}
\begin{tabular}{l cccc cc}
\toprule
 & \multicolumn{4}{c}{memory-order estimate} & \multicolumn{2}{c}{$\hat a_n$}\\
\cmidrule(lr){2-5}\cmidrule(lr){6-7}
station & $V_{st}$ & $x_{st}$ & $\tilde\varepsilon_{st}$ & $(x_{st}-\bar x_s)\tilde\varepsilon_{st}$ & part (i) & part (ii)\\
\midrule
Valentia & 0.071 & 0.104 & 0.176 & 0.085 & 0.50 & 0.50\\
Shannon & 0.150 & 0.101 & 0.414 & $-$0.039 & 0.90 & 0.50\\
Dublin & 0.179 & 0.088 & 0.268 & 0.217 & 0.64 & 0.53\\
Claremorris & 0.139 & 0.095 & 0.478 & 0.038 & 0.90 & 0.50\\
Mullingar & 0.212 & 0.083 & 0.347 & 0.188 & 0.79 & 0.50\\
Belmullet & 0.116 & 0.094 & 0.182 & 0.092 & 0.50 & 0.50\\
Malin Head & 0.205 & 0.101 & 0.390 & 0.082 & 0.88 & 0.50\\
\bottomrule
\end{tabular}
\end{table}

From the estimates $\hat d_x$, $\hat d_\varepsilon$, and $\hat d_{x\varepsilon}$ for $x_{st}$, $\tilde\varepsilon_{st}$, and $(x_{st}-\bar x_s)\tilde\varepsilon_{st}$, we compute the exponent of Corollary~\ref{cor:plugin},
\[
\hat a_n=\min\{\max(1/2,\,2\max(\hat d_n,0)+0.1),\,0.9\},
\]
in Table~\ref{tab:memory}. Part (ii), with $\hat d_n=\max(\hat d_{x\varepsilon},\hat d_x+\hat d_\varepsilon-1/2)$, yields $0.50$ at six stations and $0.53$ at Dublin, and part (i), with $\hat d_n=\hat d_\varepsilon$, the largest estimate among the score's coordinates, would raise the exponent to the cap $0.9$ at two stations and leave $n^{0.1}$, about $2.5$, effective observations.

To imitate a short record, we cut 2000--2024 into consecutive blocks of $\ell$ days, for $\ell=365$, $548$, $730$, $1096$, and $1461$, one to four years. From block $b$ of station $s$, we compute the $95\%$ interval $I_{sb}$ for $\mu_s$ or $\beta_{1,s}$, by random weighting, by the methods of Section~\ref{sec:sim-structures}, and by least squares under independent errors. For random weighting, we draw subsampling weights with $\tau=\ell^a$, at $a=1/2$ or $a=\hat a_n$, and compute $\tilde C_n$ from the block. For $\mu_s$, the regression reduces to the intercept, so the randomly weighted estimator is the weighted mean of $V_{st}$ over the block, with $\hat a_n$ from part (i) and the block's estimate for $V_{st}$, and for $\beta_{1,s}$, $\hat a_n$ comes from part (ii) and the block's estimates. Computed within the blocks, $\hat a_n$ equals the floor $1/2$ in $95\%$ of the blocks for the slope, and averages $0.56$ over the one-year blocks and $0.51$ over the four-year blocks for the mean.

We compare the intervals with the $25$-year value $\hat\theta_s$, the least-squares estimate from the $25$ years, the mean of $V_{st}$ for $\mu_s$, and report the fraction
\[
\frac{1}{7\lfloor n/\ell\rfloor}\sum_{s=1}^{7}\sum_{b=1}^{\lfloor n/\ell\rfloor}\ind\{\hat\theta_s\in I_{sb}\},
\]
averaged over $200$ draws of the weights for the randomized intervals. The $25$-year value is an estimate from data including the block, and blocks within a series are dependent, so the fraction measures agreement between a short record and the long record. To see how far agreement is from coverage, we simulate $1000$ series of fractional Gaussian noise of length $n=8946$, with memory order $0.1$ or $0.2$ and blocks of $365$ or $1461$ days. The fraction of intervals from random weighting at $a=1/2$ containing the full-sample mean is within $0.021$ of the fraction containing the true mean.

For the standard error of a fraction, we average the indicators over the seven stations within a period and divide the standard deviation of the $\lfloor n/\ell\rfloor$ period averages by $\lfloor n/\ell\rfloor^{1/2}$. The standard errors are at most $0.02$ for the randomized intervals and up to $0.11$ for SN and EWC at $\ell=1461$, and dependence between periods makes the true standard errors larger.

\subsection{Results}\label{sec:app-results}

\begin{figure}[!t]
\centering
\includegraphics[width=\textwidth]{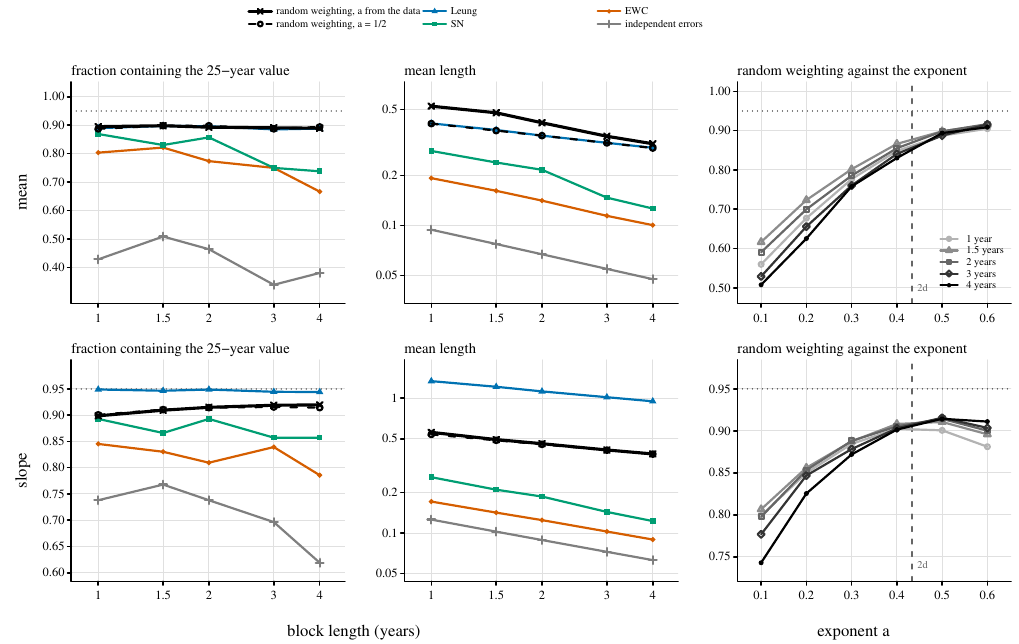}
\caption{Fraction of the $95\%$ intervals containing the $25$-year value and mean length against block length, and the fraction for random weighting against $a$, for the mean (top) and the slope (bottom).}
\label{fig:wind}
\end{figure}

In Table~\ref{tab:wind} and Figure~\ref{fig:wind}, intervals for the mean computed under independence between days contain the $25$-year value in $0.43$ of the one-year blocks, in agreement with \citet{HaslettRaftery1989}, and in $0.38$ of the four-year blocks. Under long memory, a block mean's variance is of order $\ell^{2d-1}$, against the order $\ell^{-1}$ presumed under independence. For SN and EWC, the fraction decreases from $0.87$ and $0.80$ at one year to $0.74$ and $0.67$ at four. For random weighting, the fraction is between $0.89$ and $0.90$ for the mean and between $0.90$ and $0.92$ for the slope at every block length, at $a=1/2$ and at $a=\hat a_n$.

\begin{table}[!ht]
\caption{Fraction of the $95\%$ intervals containing the $25$-year value, by block length $\ell$ in days, with $\lfloor n/\ell\rfloor$ periods of seven stations.}
\label{tab:wind}
\centering\small
\setlength{\tabcolsep}{4.5pt}
\begin{tabular}{l ccccc ccccc}
\toprule
 & \multicolumn{5}{c}{mean $\mu_s$} & \multicolumn{5}{c}{slope $\beta_{1,s}$}\\
\cmidrule(lr){2-6}\cmidrule(lr){7-11}
$\ell$ & $365$ & $548$ & $730$ & $1096$ & $1461$ & $365$ & $548$ & $730$ & $1096$ & $1461$\\
$\lfloor n/\ell\rfloor$ & $24$ & $16$ & $12$ & $8$ & $6$ & $24$ & $16$ & $12$ & $8$ & $6$\\
\midrule
random weighting, $a=\hat a_n$ & 0.89 & 0.90 & 0.89 & 0.89 & 0.89 & 0.90 & 0.91 & 0.92 & 0.92 & 0.92\\
random weighting, $a=1/2$ & 0.89 & 0.90 & 0.90 & 0.89 & 0.89 & 0.90 & 0.91 & 0.91 & 0.92 & 0.91\\
\citet{Leung2022} & 0.89 & 0.90 & 0.90 & 0.89 & 0.89 & 0.95 & 0.95 & 0.95 & 0.94 & 0.94\\
SN & 0.87 & 0.83 & 0.86 & 0.75 & 0.74 & 0.89 & 0.87 & 0.89 & 0.86 & 0.86\\
EWC & 0.80 & 0.82 & 0.77 & 0.75 & 0.67 & 0.85 & 0.83 & 0.81 & 0.84 & 0.79\\
independent errors & 0.43 & 0.51 & 0.46 & 0.34 & 0.38 & 0.74 & 0.77 & 0.74 & 0.70 & 0.62\\
\bottomrule
\end{tabular}
\end{table}

In Figure~\ref{fig:wind}, the interval of \citet{Leung2022} for the slope has about $2.5$ times the mean length of random weighting's at $a=1/2$, $1.34$ against $0.54$ at one year and $0.95$ against $0.39$ at four. For random weighting, the fraction increases with $a$ up to the dashed line at $0.43$, twice the largest estimate of Table~\ref{tab:memory} for $V_{st}$ and $(x_{st}-\bar x_s)\tilde\varepsilon_{st}$, and changes little beyond, consistent with the rate condition.

A gap to $0.95$ remains, of the size seen in Table~\ref{tab:outside} at $n=1000$ for memory orders $0.1$ and $0.2$, with effective sample sizes $\ell^{1-a}$ ranging here from $365^{0.44}=13$ to $1461^{1/2}=38$. Any drift in the mean speed over the $25$ years adds to the gap, breaks and long memory being hard to distinguish (see \citet{DieboldInoue2001}). To measure the effect of such a drift, we subtract from $V_{st}$ a linear trend fitted by least squares over 2000--2024 at every station and repeat the computations. The memory-order estimates for $V_{st}$ then change by at most $0.05$, and the fraction for random weighting rises by at most $0.034$ for the mean and changes by at most $0.001$ for the slope. The fractions for SN, EWC, and independent errors rise by up to $0.19$ for the mean.

\section{Discussion}\label{sec:discussion}

Random weighting yields confidence regions for regression coefficients from the sandwich covariance of \citet{White1980} and chi-squared critical values, under a bound on the growth rate of the score's long-run covariance. Under ergodicity, a bandwidth satisfying the bound exists, and under short memory, every $\taun\to\infty$ with $\taun/n\to0$ does. Under long memory of order $d$, the condition becomes $\taun/n^{2d}\to\infty$, met by the default $\taun=n^{1/2}$ for $d<1/4$ and by $\hat\tau_n$ of Corollary~\ref{cor:plugin}, computed from estimates of the memory orders for the score or for the slopes. To first order, the method adds to least squares a noise with covariance $(\taun/n)\Wm$, and the confidence statements hold when the noise dominates the least-squares error, whatever the limit of the least-squares error. In further simulations, reported in the Supplementary Material, Gaussian noise of the same scale had about the coverage of the weights, and the regions averaged over draws of the weights and the confidence function $C_{n,J}$ had coverages within $0.035$ of the limits of Proposition~\ref{prop:limits}.

Under long memory, random weighting's coverage with the data-driven exponent increases with $n$ in every design of Table~\ref{tab:outside}, to $0.892$ at $d=0.3$ and $n=64{,}000$, against $0.726$ for SN and $0.225$ for EWC. In the wind data, about nine tenths of the intervals from random weighting contain the $25$-year value, for records of one to four years, with a single tuning constant set by the data. In our simulations, the coverage with $\tilde C_n$ at a fixed bandwidth was within $0.035$ of the prediction once $\neff$ exceeded $40$, and between $0.886$ and $0.962$ with $\hat\tau_n$ for memory orders up to $0.2$. At equal length and $\neff$ of about $50$ or more, random weighting, of length independent of $\beta$, has about the coverage of the resampled intervals of \citet{Leung2022}, valid for $d<1/4$, and at $d=0.3$, coverage increases with $n$ for random weighting and decreases for the resampled intervals.

In practice, we recommend $\tilde C_n$, computed with unit weights, and $\hat\tau_n$ of Corollary~\ref{cor:plugin} with $a_0=1/2$, part (i) for the score and part (ii) for the slopes when the memory-order estimates of the coordinates of $\zt\tilde\varepsilon_t$ are largest for the intercept, the case of joint long memory in our simulations and of the application's slope regression. When a bound on the ratio of the score's long-run to marginal variance is available, the enlarged region of Corollary~\ref{cor:cost}(iii) keeps the level, with coverage at least $0.979$ in our designs with $d=0$, and under strong short memory in a small sample, a larger $\taun$ raises coverage, in agreement with Corollary~\ref{cor:cost}(i). Reporting the auxiliary sample's seed with the results makes the region reproducible. With more than one regressor, the case of joint long memory in our simulations, or with regressors dependent on the errors, parts (ii) and (iii) of Proposition~\ref{prop:rates} apply to the score's coordinates one by one, and uniformity of the guarantees over classes of dependence, the question of \citet{PreinerstorferPotscher2016}, is a direction for further work.

\clearpage
\appendix
\section*{Appendix}
\refstepcounter{section}\label{app:proofs}

\begin{proof}[Proof of Proposition~\ref{prop:identity}]
(i) Substituting $Y_t=\zt^\T\beta+\et$ in \eqref{eq:estimator} yields
\[
\frac1n\sum_{t=1}^n\wnt\zt Y_t=\Gn\beta+\frac1n\sum_{t=1}^n\wnt\st=\Gn\beta+A_n+\sbar ,
\]
since $\wnt=(\wnt-1)+1$, and (i) follows on multiplying by $\Gn^{-1}$ whenever $\Gn$ is nonsingular.

(ii) Since the weights are independent of the data, given $\mathcal D_n$, $w_{n1}-1,\dots,w_{nn}-1$ are independent with mean $0$ and variance $\taun-1$, and $s_1,\dots,s_n$ are constants. Hence $\E[A_n\mid\mathcal D_n]=0$ and
\[
\Cov(A_n\mid\mathcal D_n)=\frac1{n^2}\sum_{t=1}^n\Var(\wnt)\,\st\st^\T=\frac{\taun-1}{n^2}\sum_{t=1}^n\st\st^\T .
\]

(iii) As $\sbar$ is $\mathcal D_n$-measurable, $\E[A_n\sbar^\T]=\E\{\E[A_n\mid\mathcal D_n]\,\sbar^\T\}=0$ by (ii), so $\Cov(A_n,\sbar)=0$ and $\Var(A_n+\sbar)=\Var(A_n)+\Var(\sbar)$. By (ii) and stationarity,
\[
\Var(A_n)=\E\{\Cov(A_n\mid\mathcal D_n)\}=\frac{\taun-1}{n^2}\sum_{t=1}^n\E[\st\st^\T]=\frac{\taun-1}{n}\,\Vm ,
\]
and $n\Var(\sbar)=\Ln$, so $n\Var(A_n+\sbar)=(\taun-1)\Vm+\Ln$.
\end{proof}

\begin{proof}[Proof of Lemma~\ref{lem:clt}]
For fixed $\theta\in\R^m$, we set $\omega_{nt}=(n\taun)^{-1/2}(\wnt-1)\theta^\T\st$, so $\theta^\T(n/\taun)^{1/2}A_n=\sum_t\omega_{nt}$. Since the weights are independent of the data and of each other, all bounded measurable $g_1,\dots,g_n$ satisfy
\begin{equation}\label{eq:prod}
\E\Big[\prod_{t=1}^ng_t(\omega_{nt})\,\Big|\,\mathcal D_n\Big]=\prod_{t=1}^n\E g_t\{(n\taun)^{-1/2}(\wnt-1)\,a\}\Big|_{a=\theta^\T\st}.
\end{equation}
Given $\mathcal D_n$, the $\omega_{nt}$ are therefore independent, with $\E[\omega_{nt}\mid\mathcal D_n]=0$ and $\E[\omega_{nt}^2\mid\mathcal D_n]=(\taun-1)(\theta^\T\st)^2/(n\taun)$. Their conditional variances add up to
\[
\sigma_n^2=\Big(1-\frac1{\taun}\Big)\frac1n\sum_{t=1}^n(\theta^\T\st)^2 ,
\]
and $\sigma_n^2\pto(1-\lambda)\theta^\T\Vm\theta$ by (A2) and $1/\taun\to\lambda$.

For $\eta>0$ and $c>0$, we set
\[
\mathcal L_n(\eta)=\sum_{t=1}^n\E[\omega_{nt}^2\ind\{|\omega_{nt}|>\eta\}\mid\mathcal D_n],\qquad
\rho_n(c)=\frac1{\taun}\,\E\big[(w_{n1}-1)^2\ind\{|w_{n1}-1|>c\,(n\taun)^{1/2}\}\big].
\]
For $c_1>0$ and $|\theta^\T\st|\le c_1$, $\{|\omega_{nt}|>\eta\}$ is contained in $\{|\wnt-1|>\eta(n\taun)^{1/2}/c_1\}$, and for $|\theta^\T\st|>c_1$,
\[
\E[\omega_{nt}^2\ind\{|\omega_{nt}|>\eta\}\mid\mathcal D_n]\le\E[\omega_{nt}^2\mid\mathcal D_n]\le(\theta^\T\st)^2/n .
\]
Hence
\[
\mathcal L_n(\eta)\le\rho_n(\eta/c_1)\,\frac1n\sum_{t=1}^n(\theta^\T\st)^2+\frac1n\sum_{t=1}^n(\theta^\T\st)^2\ind\{|\theta^\T\st|>c_1\}.
\]
Since $\E(w_{n1}-1)^4\le8(\E w_{n1}^4+1)$,
\[
\rho_n(c)\le\frac{\E(w_{n1}-1)^4}{c^2n\taun^2}\le\frac{8(\E w_{n1}^4+1)}{c^2n\taun^2}\to0 ,
\]
so the first term of the bound is $o_{\Prob}(1)$ by (A2). By stationarity, the expectation of the second term is $\E[(\theta^\T s_1)^2\ind\{|\theta^\T s_1|>c_1\}]$, arbitrarily small for large $c_1$. Therefore $\mathcal L_n(\eta)\pto0$ for every $\eta>0$.

With $g_t(u)=e^{iu}$ for every $t$, \eqref{eq:prod} becomes
\[
\phi_n(\theta)=\E\big[\exp\{i\theta^\T(n/\taun)^{1/2}A_n\}\mid\mathcal D_n\big]=\prod_{t=1}^n\psi_n\{(n\taun)^{-1/2}\theta^\T\st\},
\]
where $\psi_n$ is the characteristic function of $w_{n1}-1$. For a fixed realization of the data, the right side is the characteristic function of a sum of independent centered variables with variances adding up to $\sigma_n^2$ and Lindeberg sum $\mathcal L_n(\eta)$. By the Lindeberg central limit theorem, $\phi_n(\theta)\to\exp\{-(1-\lambda)\theta^\T\Vm\theta/2\}$ along every subsequence on which $\sigma_n^2\to(1-\lambda)\theta^\T\Vm\theta$ and $\mathcal L_n(\eta)\to0$ for every $\eta>0$. When $(1-\lambda)\theta^\T\Vm\theta=0$, the limit $1$ follows from $\sigma_n^2\to0$.

Both conditions hold in probability, the first by the limit of $\sigma_n^2$ and the second by the bound on $\mathcal L_n(1/k)$ for every integer $k\ge1$. A diagonal argument over $k$ yields, in every subsequence, a further subsequence along which $\sigma_n^2\to(1-\lambda)\theta^\T\Vm\theta$ and $\mathcal L_n(1/k)\to0$ for all $k$ almost surely, and $\mathcal L_n(\eta)\le\mathcal L_n(1/k)$ for $\eta\ge1/k$. Along the further subsequence, $\phi_n(\theta)$ converges to $\exp\{-(1-\lambda)\theta^\T\Vm\theta/2\}$ for almost every realization of the data. As the initial subsequence is arbitrary, $\phi_n(\theta)\pto\exp\{-(1-\lambda)\theta^\T\Vm\theta/2\}$.

For the conditional form, we substitute $\hat\tau_n$ for $\taun$ in $\omega_{nt}$, $\sigma_n^2$, and $\rho_n$. Given $\mathcal D_n$, the weights are independent with mean $1$ and second moment $\hat\tau_n$, and every expectation over the weights above, the right side of \eqref{eq:prod} included, becomes a conditional expectation given $\mathcal D_n$, and the bound on $\rho_n$ becomes
\[
\rho_n(c)\le\frac{8(\E[w_{n1}^4\mid\mathcal D_n]+1)}{c^2n\hat\tau_n^2}\pto0 .
\]
Since $1/\hat\tau_n\pto\lambda$, $\sigma_n^2\pto(1-\lambda)\theta^\T\Vm\theta$ as well, and the diagonal argument and the Lindeberg central limit theorem for a fixed realization of the data yield the limit of $\phi_n(\theta)$ as before.
\end{proof}

\begin{proof}[Proof of Corollary~\ref{cor:random}]
(i) With $c_1>0$ and $\Delta_n=\Gn-\hat\Qm_n=\Delta_{n,1}+\Delta_{n,2}$ as in the proof of Theorem~\ref{thm:limit}(i), given $\mathcal D_n$, the weights are independent with mean $1$ and variance $\hat\tau_n-1$, and $\E[w_{n1}^4\mid\mathcal D_n]\ge\hat\tau_n^3$, so
\[
\E\big[\|\Delta_{n,1}\|^2\mid\mathcal D_n\big]=\frac{\hat\tau_n-1}{n^2}\sum_{t=1}^n\|\zt\|^4\ind\{\|\zt\|\le c_1\}\le\frac{c_1^4\,\E[w_{n1}^4\mid\mathcal D_n]}{n\hat\tau_n^2}\pto0 .
\]
Hence, for every $\delta>0$,
\[
\Prob(\|\Delta_{n,1}\|>\delta)\le\E\min\{1,\E[\|\Delta_{n,1}\|^2\mid\mathcal D_n]/\delta^2\}\to0,
\]
and the bound on $\Delta_{n,2}$ holds with the conditional mean $1$ of the weights. Thus $\Gn\pto\Qm$ and $\Prob(E_n)\to1$.

(ii) By the conditional form of Lemma~\ref{lem:clt} with $\lambda=0$, for every $\theta\in\R^m$,
\[
\E\big[\exp\{i\theta^\T(n/\hat\tau_n)^{1/2}A_n\}\mid\mathcal D_n\big]\pto\exp(-\theta^\T\Vm\theta/2),
\]
and $(n/\hat\tau_n)^{1/2}A_n\dto N(0,\Vm)$ follows as in the proof of Theorem~\ref{thm:limit}(ii).

(iii) With $T_n=(n/\hat\tau_n)^{1/2}A_n$ and $U_n=(n/\hat\tau_n)^{1/2}\sbar$, the proof of Theorem~\ref{thm:limit}(iii) applies without change, $U_n$ being $\mathcal D_n$-measurable. For the last assertion, $(n/\hat\tau_n)^{1/2}\|\sbar\|\le(n/\tau_n^-)^{1/2}\|\sbar\|$ on $\{\hat\tau_n\ge\tau_n^-\}$, an event of asymptotic probability one, so \eqref{eq:D} for $\tau_n^-$ implies \eqref{eq:D} for $\hat\tau_n$.
\end{proof}

\begin{proof}[Proof of Lemma~\ref{lem:sandwich}]
Write $\delta_n=\bhat-\beta$. Since $\E w_{n1}^4\ge\taun^3$, the hypothesis on the weights implies $\taun/n\to0$, so $\Gn\pto\Qm$ and $\Prob(E_n)\to1$ by the argument of Theorem~\ref{thm:limit}(i). By Proposition~\ref{prop:identity}(ii) and stationarity, $\E\|A_n\|^2=(\taun-1)\E\|s_1\|^2/n\to0$, and $\delta_n=\Gn^{-1}(A_n+\sbar)$ on $E_n$ by Proposition~\ref{prop:identity}(i), so $\delta_n\pto0$.

Substituting $\ehat=\et-\zt^\T\delta_n$ in $\hat\Vm_n$ yields
\[
\hat\Vm_n=\frac1n\sum_{t=1}^n\frac{\wnt^2}{\taun}\st\st^\T-\frac2n\sum_{t=1}^n\frac{\wnt^2}{\taun}\zt\zt^\T\et\,\zt^\T\delta_n+\frac1n\sum_{t=1}^n\frac{\wnt^2}{\taun}\zt\zt^\T(\zt^\T\delta_n)^2 .
\]
In norm, the second and the third term are at most
\[
2\|\delta_n\|\,\frac1n\sum_{t=1}^n\frac{\wnt^2}{\taun}\|\zt\|^3|\et|\qquad\text{and}\qquad\|\delta_n\|^2\,\frac1n\sum_{t=1}^n\frac{\wnt^2}{\taun}\|\zt\|^4 .
\]
As $\E[\wnt^2/\taun]=1$ and the weights are independent of the data, the two sums have expectations $\E[\|z_1\|^3|\varepsilon_1|]$ and $\E\|z_1\|^4$, finite under (A1) and (A4) because $\E[\|z_1\|^3|\varepsilon_1|]\le(\E\|z_1\|^4)^{1/2}(\E\|s_1\|^2)^{1/2}$. Both sums are therefore $O_{\Prob}(1)$, and the two terms are $o_{\Prob}(1)$.

It remains to prove $\Gamma_n=n^{-1}\sum_t(\wnt^2/\taun-1)\st\st^\T\pto0$, since the first term is $n^{-1}\sum_t\st\st^\T+\Gamma_n$ and $n^{-1}\sum_t\st\st^\T\pto\Vm$ by (A2). For $c_1>0$, we split $\Gamma_n=\Gamma_{n,1}+\Gamma_{n,2}$ with the indicators $\ind\{\|\st\|\le c_1\}$ and $\ind\{\|\st\|>c_1\}$. Given $\mathcal D_n$, $\Gamma_{n,1}$ is a sum of independent centered terms, and $\Var(\wnt^2)\le\E w_{n1}^4$, so
\[
\E[\|\Gamma_{n,1}\|^2\mid\mathcal D_n]=\frac{\Var(\wnt^2)}{\taun^2n^2}\sum_{t=1}^n\|\st\|^4\ind\{\|\st\|\le c_1\}\le\frac{c_1^4\,\E w_{n1}^4}{n\taun^2}\to0 .
\]
For the other piece, $\|\Gamma_{n,2}\|\le n^{-1}\sum_t(\wnt^2/\taun+1)\|\st\|^2\ind\{\|\st\|>c_1\}$ has expectation $2\E[\|s_1\|^2\ind\{\|s_1\|>c_1\}]$, arbitrarily small for large $c_1$. Hence $\Gamma_n\pto0$, as in the proof of Theorem~\ref{thm:limit}(i).
\end{proof}

\begin{proof}[Proof of Corollary~\ref{cor:slopes}]
Under ergodicity, (A2) holds and $\bar\varepsilon_n\to0$ almost surely, so Theorem~\ref{thm:limit}(i) and (ii) apply. With $X_t^c=X_t-\mu_X$, $\tilde X_t=X_t-\bar X_{n,w}$, and $P=[-\mu_X,I_p]\in\R^{p\times m}$, we have $P\zt=X_t^c$, $P\st=X_t^c\et$, $P\Vm P^\T=\Vm_X$, and $e_1^\T\st=\et$. Whenever $\Gn$ is nonsingular, eliminating the intercept from the weighted normal equations and substituting $Y_t=\beta_0+X_t^\T\beta_X+\et$, with $\sum_t\wnt\tilde X_t=0$, we obtain
\[
\hat\beta^{w}_{n,X}-\beta_X=\Big(\frac1n\sum_{t=1}^n\wnt\tilde X_t\tilde X_t^\T\Big)^{-1}\frac1n\sum_{t=1}^n\wnt\tilde X_t\et .
\]

Since $n^{-1}\sum_t\wnt$, $n^{-1}\sum_t\wnt X_t$, and $n^{-1}\sum_t\wnt X_tX_t^\T$ are blocks of $\Gn$, Theorem~\ref{thm:limit}(i) yields
\[
\frac1n\sum_{t=1}^n\wnt\tilde X_t\tilde X_t^\T=\frac1n\sum_{t=1}^n\wnt X_tX_t^\T-\Big(\frac1n\sum_{t=1}^n\wnt\Big)\bar X_{n,w}\bar X_{n,w}^\T\pto\E[X_1X_1^\T]-\mu_X\mu_X^\T=\Sigma_X .
\]
Writing $\tilde X_t=X_t^c-(\bar X_n-\mu_X)-(\bar X_{n,w}-\bar X_n)$ and $n^{-1}\sum_t\wnt\st=A_n+\sbar$, we obtain
\begin{align*}
(n/\taun)^{1/2}\frac1n\sum_{t=1}^n\wnt\tilde X_t\et
&=P(n/\taun)^{1/2}A_n+(n/\taun)^{1/2}\frac1n\sum_{t=1}^n(X_t-\bar X_n)\et\\
&\quad-(\bar X_n-\mu_X)(n/\taun)^{1/2}e_1^\T A_n-(\bar X_{n,w}-\bar X_n)(n/\taun)^{1/2}e_1^\T(A_n+\sbar).
\end{align*}

Of the four terms, the first converges in distribution to $N(0,\Vm_X)$ by Theorem~\ref{thm:limit}(ii), the second is $o_{\Prob}(1)$ by \eqref{eq:DX}, and the third is $o_{\Prob}(1)$ because $\bar X_n\pto\mu_X$ by (A2) and $(n/\taun)^{1/2}A_n=O_{\Prob}(1)$. In the fourth,
\[
\bar X_{n,w}-\bar X_n=\frac{n^{-1}\sum_{t=1}^n(\wnt-1)(X_t-\bar X_n)}{n^{-1}\sum_{t=1}^n\wnt}
\]
has a numerator with conditional mean $0$ and conditional second moment $(\taun-1)n^{-2}\sum_t\|X_t-\bar X_n\|^2=O_{\Prob}(\taun/n)$, so $\bar X_{n,w}-\bar X_n=O_{\Prob}((\taun/n)^{1/2})$. Since the other factor is the sum of $(n/\taun)^{1/2}e_1^\T A_n=O_{\Prob}(1)$ and $(n/\taun)^{1/2}\bar\varepsilon_n$, the fourth term is $o_{\Prob}(1)+O_{\Prob}(1)\,\bar\varepsilon_n\pto0$. Therefore
\[
(n/\taun)^{1/2}(\hat\beta^{w}_{n,X}-\beta_X)\dto\Sigma_X^{-1}N(0,\Vm_X)=N(0,\Wm_X).
\]

If (A4) holds as well, $\sbar\to0$ almost surely by ergodicity, so $(n/\taun)\Chat\pto\Wm$ by Lemma~\ref{lem:sandwich} and Theorem~\ref{thm:limit}(i), and the slope block of $(n/\taun)\Chat$ converges to $\Wm_X$, a positive definite matrix. Then the Wald statistic of $\beta_X$, the restricted statistics, and the marginal ratios converge as in the proof of Theorem~\ref{thm:feasible}.
\end{proof}

\begin{proof}[Proof of Proposition~\ref{prop:rates}]
(i) With $u_t=v^\T\st$ and $\gamma(k)=\Cov(u_1,u_{1+k})$,
\[
n\Var(\bar u_n)=\gamma(0)+2\sum_{k=1}^{n-1}\Big(1-\frac kn\Big)\gamma(k).
\]
Under short memory, the right side is at most $C=\gamma(0)+2\sum_{k\ge1}|\gamma(k)|$, so $(n/\taun)\E(v^\T\sbar)^2\le C/\taun\to0$.

(ii) Under long memory of order $d$ with constant $c$, $\gamma(k)=c\,k^{2d-1}\{1+o(1)\}$, and
\[
2\sum_{k=1}^{n-1}\Big(1-\frac kn\Big)k^{2d-1}=\frac{2n^{2d}}{n}\sum_{k=1}^{n-1}\Big(1-\frac kn\Big)\Big(\frac kn\Big)^{2d-1}\sim2n^{2d}\!\int_0^1(1-x)x^{2d-1}dx=\frac{n^{2d}}{d(2d+1)},
\]
the Riemann sums converging because $(1-x)x^{2d-1}$ is decreasing and integrable on $(0,1)$. By the same computation, the $o(1)$ terms contribute $o(n^{2d})$, and $\gamma(0)=o(n^{2d})$. Hence $\Var(\sqrt n\,v^\T\sbar)\sim c\,n^{2d}/\{d(2d+1)\}$, and $(n/\taun)\E(v^\T\sbar)^2\sim c\,n^{2d}/\{d(2d+1)\taun\}\to0$ when $\taun/n^{2d}\to\infty$.

(iii) If $\taun/n^{2d}$ is bounded along a subsequence, then $n^d/\taun^{1/2}\ge c_0$ along the same subsequence for some $c_0>0$, and, since $(n/\taun)^{1/2}v^\T\sbar=(n^d/\taun^{1/2})\,n^{1/2-d}v^\T\sbar$,
\[
\Prob\{|(n/\taun)^{1/2}v^\T\sbar|>c_0\eta\}\ge\Prob(|n^{1/2-d}v^\T\sbar|>\eta)\to\Prob(|Y|>\eta)
\]
along the subsequence, at every continuity point $\eta$ of the distribution of $|Y|$. Since $\Prob(Y\ne0)>0$, the limit is positive for some continuity point $\eta>0$, so $(n/\taun)^{1/2}v^\T\sbar$ is not $o_{\Prob}(1)$.

(iv) With $u_t=(X_t-\mu_X)\et$, the independence of the two sequences and $\E\varepsilon_1=0$ yield
\[
\E u_t=0,\qquad \E u_t^2=\Var(X_1)\Var(\varepsilon_1)<\infty,\qquad \Cov(u_1,u_{1+k})=\gamma_X(k)\gamma_\varepsilon(k),
\]
where $\gamma_X$ and $\gamma_\varepsilon$ are the autocovariances of the two sequences. If $d_X+d_\varepsilon>1/2$, both sequences have long memory, with constants $c_X$ and $c_\varepsilon$, and $\gamma_X(k)\gamma_\varepsilon(k)\sim c_Xc_\varepsilon k^{2(d_X+d_\varepsilon-1/2)-1}$, so $(u_t)$ has long memory of order $d_X+d_\varepsilon-1/2\in(0,1/2)$. If $d_X+d_\varepsilon<1/2$ and one sequence has short memory, then $\sum_k|\gamma_X(k)\gamma_\varepsilon(k)|<\infty$, because the other sequence has bounded autocovariances. If both have long memory, then $\gamma_X(k)\gamma_\varepsilon(k)\sim c_Xc_\varepsilon k^{2(d_X+d_\varepsilon)-2}$ with $2(d_X+d_\varepsilon)-2<-1$, and the sum is again finite. If $d_X+d_\varepsilon=1/2$, both sequences have long memory and $\gamma_X(k)\gamma_\varepsilon(k)\sim c_Xc_\varepsilon k^{-1}$, so, as in (ii), $n\Var(\bar u_n)\sim2c_Xc_\varepsilon\log n$, and $(n/\taun)\E\bar u_n^2\to0$ when $\taun/\log n\to\infty$.

For \eqref{eq:DX}, we write
\[
\frac1n\sum_{t=1}^n(X_t-\bar X_n)\et=\bar u_n-(\bar X_n-\mu_X)\bar\varepsilon_n .
\]
As $u_t=v^\T\st$ for $v=(-\mu_X,1)^\T$, (i), (ii), and the case $d_X+d_\varepsilon=1/2$ above yield $(n/\taun)^{1/2}\bar u_n\pto0$ under the condition of each case. For the cross term, by independence again,
\[
(n/\taun)\,\E\{(\bar X_n-\mu_X)^2\bar\varepsilon_n^2\}=\frac{\Var(\sqrt n\,\bar X_n)\Var(\sqrt n\,\bar\varepsilon_n)}{n\taun},
\]
and each variance is $O(n^{2d})$ for the corresponding order $d$ by the computations of (i) and (ii). Therefore the left side is $O(n^{2(d_X+d_\varepsilon)-1}/\taun)=o(1)$ under the condition of each case, which establishes \eqref{eq:DX}.
\end{proof}

\begin{proof}[Proof of Lemma~\ref{lem:shift}]
As in the proof of Lemma~\ref{lem:sandwich}, $\taun/n\to0$, $\Gn\pto\Qm$, and $\Prob(E_n)\to1$. Since $n/\taun\to\infty$ and $(n/\taun)^{1/2}\sbar=O_{\Prob}(1)$, $\sbar\pto0$, so Lemma~\ref{lem:sandwich} yields $\hat\Vm_n\pto\Vm$, hence $\Prob(F_n)\to1$ and $M_n=(n/\taun)\Chat\pto\Wm$, as $M_n=\Gn^{-1}\hat\Vm_n\Gn^{-1}$ on $F_n$. For fixed $x>0$ and $Z_n=(n/\taun)^{1/2}(\bhat-\bols)$,
\[
(\bhat-\beta)^\T\Chat^{-1}(\bhat-\beta)=(Z_n+\Omega_n)^\T M_n^{-1}(Z_n+\Omega_n).
\]
Here $\Omega_n$ is $\mathcal D_n$-measurable, and $\Omega_n=\hat\Qm_n^{-1}(n/\taun)^{1/2}\sbar$ whenever $\hat\Qm_n$ is nonsingular, an event of asymptotic probability one by (A2), so $\Omega_n=O_{\Prob}(1)$.

Let $Z\sim N(0,(1-\lambda)\Wm)$, and let $f$ be bounded and Lipschitz. On $E_n\cap\{\hat\Qm_n\ \text{nonsingular}\}$, Proposition~\ref{prop:identity}(i) yields $\bhat-\bols=\Gn^{-1}A_n+(\Gn^{-1}-\hat\Qm_n^{-1})\sbar$, so
\[
Z_n=\Qm^{-1}(n/\taun)^{1/2}A_n+\varrho_n,\qquad \varrho_n\pto0 ,
\]
because $(n/\taun)^{1/2}A_n=O_{\Prob}(1)$ by Proposition~\ref{prop:identity}(ii), $(n/\taun)^{1/2}\sbar=O_{\Prob}(1)$, $\Gn\pto\Qm$, and $\hat\Qm_n\pto\Qm$. Then $\E|f(Z_n)-f(\Qm^{-1}(n/\taun)^{1/2}A_n)|\to0$, and the same holds for the conditional expectations given $\mathcal D_n$, in $L^1$.

By Lemma~\ref{lem:clt}, the conditional characteristic function of $(n/\taun)^{1/2}A_n$ converges in probability at every $\theta$ to the characteristic function of $N(0,(1-\lambda)\Vm)$. Every subsequence has a further subsequence along which the convergence holds almost surely for all rational $\theta$ and $\E[\|(n/\taun)^{1/2}A_n\|^2\mid\mathcal D_n]=(1-1/\taun)n^{-1}\sum_t\|\st\|^2$ converges. Along the further subsequence, the conditional characteristic functions are Lipschitz in $\theta$ with a bounded constant, hence converge for every $\theta$, and the conditional distributions of $(n/\taun)^{1/2}A_n$ converge weakly to $N(0,(1-\lambda)\Vm)$ almost surely. As the initial subsequence is arbitrary, $\E[f(\Qm^{-1}(n/\taun)^{1/2}A_n)\mid\mathcal D_n]\pto\E f(Z)$, and hence $\E[f(Z_n)\mid\mathcal D_n]\pto\E f(Z)$.

With $q_b(z)=(z+b)^\T\Wm^{-1}(z+b)$ for $b\in\R^m$, $Y_n'=q_{\Omega_n}(Z_n)$, $Y_n=(Z_n+\Omega_n)^\T M_n^{-1}(Z_n+\Omega_n)$, and $\zeta=(1-\lambda)^{-1/2}\Wm^{-1/2}Z$ standard normal,
\[
q_b(Z)=(1-\lambda)\|\zeta+(1-\lambda)^{-1/2}\Wm^{-1/2}b\|^2,
\]
so $\Prob\{q_b(Z)\le x\}=\Pi_x(b)$. For $c>1$, the convergence of the conditional expectations applied to $f(z)=\min\{1,\operatorname{dist}(z,\{u\mid\|u\|\le c-1\})\}$ yields $\limsup_n\Prob(\|Z_n\|>c)\le\Prob(\|Z\|>c-1)$, so $Z_n=O_{\Prob}(1)$ and
\[
Y_n-Y_n'=(Z_n+\Omega_n)^\T(M_n^{-1}-\Wm^{-1})(Z_n+\Omega_n)\pto0.
\]

For $c>0$, $0<\delta<x/4$, $\psi(u)=\min\{1,\max(u,0)\}$, and $\|b\|\le c$, the functions
\[
g_b^-(z)=\psi\Big\{\frac{x-q_b(z)}{\delta}\Big\}\qquad\text{and}\qquad g_b^+(z)=\psi\Big\{\frac{x+\delta-q_b(z)}{\delta}\Big\}
\]
satisfy $g_b^-\le\ind\{q_b\le x\}\le g_b^+$. Since $q_b$ has a bounded gradient on the bounded set where $g_b^\pm$ varies, the family $\mathcal G=\{g_b^\pm\mid\|b\|\le c\}$ is uniformly bounded and uniformly Lipschitz. A diagonal argument over a countable convergence-determining family of bounded Lipschitz functions yields, in every subsequence, a further subsequence along which the conditional distributions of $Z_n$ converge weakly to the distribution of $Z$ almost surely. Weak convergence is convergence in the bounded Lipschitz metric (see \citet[Theorem~11.3.3]{Dudley2002}), and the bounded Lipschitz norms of the members of $\mathcal G$ are bounded, so the convergence is uniform over $\mathcal G$ and $\sup_{g\in\mathcal G}|\E[g(Z_n)\mid\mathcal D_n]-\E g(Z)|\pto0$.

On $[x/4,3x/2]$, the distribution of $q_b(Z)$ has a density bounded by a constant $C_x$ independent of $b$. By the preceding formula, $q_b(Z)=(1-\lambda)\sum_{j=1}^m(\zeta_j+\mu_j)^2$ with $\mu=(1-\lambda)^{-1/2}\Wm^{-1/2}b$, and for a real $\mu_1$ the density of $(\zeta_1+\mu_1)^2$ at $y>0$ is
\[
\frac{\varphi(\sqrt y-\mu_1)+\varphi(\sqrt y+\mu_1)}{2\sqrt y}\le(2\pi y)^{-1/2},
\]
where $\varphi$ is the standard normal density. For $m=1$, the bound is finite on $[x/4,3x/2]$, and for $m\ge2$, one convolution of two such densities yields a bound on all of $(0,\infty)$. Therefore $|\E g_b^\pm(Z)-\Pi_x(b)|\le\Prob\{|q_b(Z)-x|\le\delta\}\le2C_x\delta$. Since $\Omega_n$ is $\mathcal D_n$-measurable, on $\{\|\Omega_n\|\le c\}$,
\[
\E[g^-_{\Omega_n}(Z_n)\mid\mathcal D_n]\le\Prob(Y_n'\le x\mid\mathcal D_n)\le\E[g^+_{\Omega_n}(Z_n)\mid\mathcal D_n],
\]
so
\[
\limsup_n\Prob\{|\Prob(Y_n'\le x\mid\mathcal D_n)-\Pi_x(\Omega_n)|>3C_x\delta\}\le\sup_n\Prob(\|\Omega_n\|>c).
\]
Since $c$ and then $\delta$ are arbitrary, $\Prob(Y_n'\le x\mid\mathcal D_n)-\Pi_x(\Omega_n)\pto0$.

It remains to pass from $Y_n'$ to $Y_n$. For the two variables, the conditional probabilities differ by at most
\[
\Prob(|Y_n-Y_n'|>\delta\mid\mathcal D_n)+\Prob(|Y_n'-x|\le\delta\mid\mathcal D_n).
\]
Here the first term converges to $0$ in $L^1$, and the second is at most
\[
\Prob(Y_n'\le x+\delta\mid\mathcal D_n)-\Prob(Y_n'\le x-2\delta\mid\mathcal D_n)\le\Pi_{x+\delta}(\Omega_n)-\Pi_{x-2\delta}(\Omega_n)+o_{\Prob}(1)\le3C_x\delta+o_{\Prob}(1),
\]
by the preceding result at the levels $x+\delta$ and $x-2\delta$. As $\delta$ is arbitrary, $\Prob(Y_n\le x\mid\mathcal D_n)-\Pi_x(\Omega_n)\pto0$.
\end{proof}

\begin{proof}[Proof of Proposition~\ref{prop:fixedtau}]
(i) As in the proof of Theorem~\ref{thm:condcov}(i), $\Gn\pto\Qm$ and $\Prob(E_n)\to1$, and $\sqrt n(\bhat-\beta)=\Gn^{-1}\sqrt n(A_n+\sbar)$ on $E_n$ by Proposition~\ref{prop:identity}(i). Since $\E w_1^4<\infty$, Lemma~\ref{lem:clt} applies with $\taun=\tau$ and $\lambda=1/\tau$, and, as $\sqrt n\,A_n=\tau^{1/2}(n/\tau)^{1/2}A_n$, for every $\theta\in\R^m$
\[
\varphi_n(\theta)=\E[\exp(i\sqrt n\,\theta^\T A_n)\mid\mathcal D_n]\pto\exp\{-(\tau-1)\theta^\T\Vm\theta/2\}.
\]
Since $\sbar$ is $\mathcal D_n$-measurable and $|\varphi_n(\theta)|\le1$,
\[
\E\exp\{i\sqrt n\,\theta^\T(A_n+\sbar)\}=\E\big[\exp(i\sqrt n\,\theta^\T\sbar)\,\varphi_n(\theta)\big]
=\exp\{-(\tau-1)\theta^\T\Vm\theta/2\}\,\E\exp(i\sqrt n\,\theta^\T\sbar)+o(1),
\]
and the right side converges to $\exp[-\theta^\T\{(\tau-1)\Vm+\Lm\}\theta/2]$ by (A7). Hence $\sqrt n(A_n+\sbar)\dto N(0,(\tau-1)\Vm+\Lm)$, and $\Gn\pto\Qm$ yields the limit of $\sqrt n(\bhat-\beta)$.

(ii) Since (A7) implies (A6), $n\Chat\pto\tau\Wm$ by Theorem~\ref{thm:condcov}(i), and by (i), with $\Sigma=\Qm^{-1}\{(\tau-1)\Vm+\Lm\}\Qm^{-1}$ and $Z\sim N(0,\Sigma)$,
\[
\{\sqrt n(\bhat-\beta)\}^\T(n\Chat)^{-1}\{\sqrt n(\bhat-\beta)\}\dto Z^\T(\tau\Wm)^{-1}Z .
\]
Here $Z^\T(\tau\Wm)^{-1}Z$ has the distribution of $\sum_ja_j\zeta_j^2$ with $a_1,\dots,a_m$ the eigenvalues of
\[
\Sigma(\tau\Wm)^{-1}=\tau^{-1}\Qm^{-1}\{(\tau-1)I_m+\Lm\Vm^{-1}\}\Qm ,
\]
and $\Lm\Vm^{-1}$ has the eigenvalues $\kappa_1,\dots,\kappa_m$ of $\Vm^{-1}\Lm$, so $a_j=(\tau-1+\kappa_j)/\tau$.

(iii) By (i), $c^\T\sqrt n(\bhat-\beta)\dto N(0,c^\T\Sigma c)$ with
\[
c^\T\Sigma c=v^\T\{(\tau-1)\Vm+\Lm\}v=(\tau-1+\kappa(v))\,v^\T\Vm v,
\]
and $n\,c^\T\Chat c\pto\tau c^\T\Wm c=\tau v^\T\Vm v>0$ by Theorem~\ref{thm:condcov}(i). Hence
\[
c^\T(\bhat-\beta)/(c^\T\Chat c)^{1/2}\dto N(0,(\tau-1+\kappa(v))/\tau).
\]
\end{proof}

\begin{proof}[Proof of Corollary~\ref{cor:cost}]
(i) Since $a_{\min}\le a_j\le a_{\max}$ and every $a_j>0$, the limit $\sum_ja_j\zeta_j^2$ in Proposition~\ref{prop:fixedtau}(ii) lies between $a_{\min}\sum_j\zeta_j^2$ and $a_{\max}\sum_j\zeta_j^2$ and has a continuous distribution. Hence $\Prob\{\beta\in R_n(\alpha)\}=\Prob\{(\bhat-\beta)^\T\Chat^{-1}(\bhat-\beta)\le\chi^2_{m,1-\alpha}\}$ converges to $\Prob(\sum_ja_j\zeta_j^2\le\chi^2_{m,1-\alpha})$, a number in the stated interval. For the marginal interval, Proposition~\ref{prop:fixedtau}(iii) with $c=e_j$ yields
\[
(\hat\beta^{w}_{n,j}-\beta_j)/(\Chat)_{jj}^{1/2}\dto N(0,(\tau-1+\kappa(v_j))/\tau),
\]
so the coverage converges to $\Prob[|\zeta_1|\le z_{1-\alpha/2}\{\tau/(\tau-1+\kappa(v_j))\}^{1/2}]$.

(ii) Whenever $\hat\Qm_n$ is nonsingular, $\sqrt n(\bols-\beta)=\hat\Qm_n^{-1}\sqrt n\,\sbar$, so $\sqrt n(\bols-\beta)\dto N(0,\Qm^{-1}\Lm\Qm^{-1})$ by (A2) and (A7), and
\[
e_j^\T\Qm^{-1}\Lm\Qm^{-1}e_j=v_j^\T\Lm v_j=\kappa(v_j)\,v_j^\T\Vm v_j.
\]

(iii) From $\kappa_1\le\kbar$ and $\tau\ge(\kbar-1)/\rhostar$, $a_{\max}=1+(\kappa_1-1)/\tau\le1+\rhostar$ and $\neff=n/\tau\le\rhostar n/(\kbar-1)$. By Proposition~\ref{prop:fixedtau}(ii) and the continuity of the distribution of $\sum_ja_j\zeta_j^2$, the coverage of the region in (iii) converges to
\[
\Prob\Big\{\sum_ja_j\zeta_j^2\le(1+\rhostar)\chi^2_{m,1-\alpha}\Big\}\ge\Prob\Big\{a_{\max}\sum_j\zeta_j^2\le(1+\rhostar)\chi^2_{m,1-\alpha}\Big\}\ge1-\alpha.
\]
\end{proof}

\begin{proof}[Proof of Proposition~\ref{prop:limits}]
We write $G_\ell$ and $A^{(\ell)}$ for $\Gn$ and $A_n$ computed from the draw $w^{(\ell)}$, and, by linearity, $\bar G=B^{-1}\sum_\ell G_\ell$ and $\bar A=B^{-1}\sum_\ell A^{(\ell)}$ are $\Gn$ and $A_n$ computed from $\bar w$. Being independent across $t$, nonnegative, with mean $1$ and variance $(\taun-1)/B$, the averaged weights form a random weighting with second moment $\tau_B$. Since $\E\bar w_1^4\le\E w_{n1}^4$ and $\tau_B\ge\taun/B$, they satisfy (A3) when the $w^{(\ell)}$ do, and (A5) with second moment $\tau_B$ when the $w^{(\ell)}$ satisfy (A5). Whenever all $G_\ell$ are nonsingular, Proposition~\ref{prop:identity}(i) for each draw and for $\bar w$ yields
\begin{equation}\label{eq:avgdiff}
\bar\beta_{n,B}-\hat\beta^{\bar w}_n=\frac1B\sum_{\ell=1}^B(G_\ell^{-1}-\bar G^{-1})(A^{(\ell)}+\sbar).
\end{equation}
For $\bar R_{n,B}(\alpha)$, $\{\beta\in\bar R_{n,B}(\alpha)\}=\{\bar S_n\le x\}$ with $\bar S_n=(\bar\beta_{n,B}-\beta)^\T(\hat C^{\bar w}_n)^{-1}(\bar\beta_{n,B}-\beta)$.

(i) For $R_n^{\mathrm{ols}}(\alpha)$, under the hypotheses of Theorem~\ref{thm:feasible},
\[
(n/\taun)^{1/2}(\bols-\beta)=\hat\Qm_n^{-1}(n/\taun)^{1/2}\sbar=o_{\Prob}(1)\qquad\text{and}\qquad(n/\taun)\Chat\pto\Wm,
\]
so $(\bols-\beta)^\T\Chat^{-1}(\bols-\beta)\pto0$ and $\Prob\{\beta\in R_n^{\mathrm{ols}}(\alpha)\}\to1$.

For $C_{n,\infty}$, Lemma~\ref{lem:shift} with $\lambda=0$ yields $p_n(\beta)-\Pi_{x_\nu}(\Omega_n)\pto0$, and $\Omega_n\pto0$ by \eqref{eq:D}, so $p_n(\beta)\pto\Pi_{x_\nu}(0)=1-\alpha+\nu>1-\alpha$ and $\Prob(\beta\in C_{n,\infty})\to1$.

For $\bar R_{n,B}(\alpha)$, in \eqref{eq:avgdiff}, $G_\ell^{-1}-\bar G^{-1}\pto0$ by Theorem~\ref{thm:limit}(i) applied to $w^{(\ell)}$ and to $\bar w$, and $(n/\taun)^{1/2}(A^{(\ell)}+\sbar)=O_{\Prob}(1)$ by Theorem~\ref{thm:limit}(ii) and \eqref{eq:D}, so $\bar\beta_{n,B}-\hat\beta^{\bar w}_n=o_{\Prob}((\taun/n)^{1/2})$. Since $\tau_B\ge\taun/B$, \eqref{eq:D} holds for $\tau_B$, and Theorem~\ref{thm:feasible} applied to $\bar w$ yields $(n/\tau_B)\hat C^{\bar w}_n\pto\Wm$ and $S_n(\beta;\bar w)\dto\chi^2_m$. As the two centers differ by $o_{\Prob}((\tau_B/n)^{1/2})$ and $(n/\tau_B)^{1/2}(\hat\beta^{\bar w}_n-\beta)=O_{\Prob}(1)$, $\bar S_n-S_n(\beta;\bar w)\pto0$, so $\bar S_n\dto\chi^2_m$ and $\Prob\{\beta\in\bar R_{n,B}(\alpha)\}\to1-\alpha$.

(ii) For $R_n^{\mathrm{ols}}(\alpha)$, under the hypotheses of Theorem~\ref{thm:condcov}, $M_n=(n/\tau)\Chat\pto\Wm$ as in the proof of Lemma~\ref{lem:shift}, and $\Omega_n=O_{\Prob}(1)$ by (A6). By the computation in the proof of Theorem~\ref{thm:condcov}(ii),
\[
(\bols-\beta)^\T\Chat^{-1}(\bols-\beta)=\Omega_n^\T M_n^{-1}\Omega_n=\Omega_n^\T\Wm^{-1}\Omega_n+o_{\Prob}(1)=\frac{\tau-1}{\tau}\Lambda_n+o_{\Prob}(1).
\]

In the case of $C_{n,\infty}$, $p_n(\beta)$ is the conditional probability of Theorem~\ref{thm:condcov}(ii) at $x_\nu$, so $p_n(\beta)-\Psi_{x_\nu}(\Lambda_n)\pto0$.

For $\bar R_{n,B}(\alpha)$, in \eqref{eq:avgdiff}, $G_\ell^{-1}-\bar G^{-1}\pto0$ by the argument of Theorem~\ref{thm:limit}(i), as $\tau/n\to0$, and $\sqrt n(A^{(\ell)}+\sbar)=O_{\Prob}(1)$ by Proposition~\ref{prop:identity}(ii) and (A6), so $\bar\beta_{n,B}-\hat\beta^{\bar w}_n=o_{\Prob}(n^{-1/2})$. Lemma~\ref{lem:shift} applies to $\bar w$ with $\lambda=1/\tau_B$, the argument $(n/\tau_B)^{1/2}(\bols-\beta)$ of $\Pi_x$, and the noncentrality
\[
\frac{n(\bols-\beta)^\T\Wm^{-1}(\bols-\beta)}{\tau_B-1}=B\Lambda_n+o_{\Prob}(1),
\]
again by the computation in the proof of Theorem~\ref{thm:condcov}(ii). As $u\mapsto\Prob\{\chi^2_m(u)\le y\}$ is continuous,
\[
\Prob\{S_n(\beta;\bar w)\le x\mid\mathcal D_n\}-\Prob\{\chi^2_m(B\Lambda_n)\le\tau_Bx/(\tau_B-1)\}\pto0.
\] Since $\sqrt n(\hat\beta^{\bar w}_n-\beta)=O_{\Prob}(1)$ and $n\hat C^{\bar w}_n\pto\tau_B\Wm$ by Theorem~\ref{thm:condcov}(i) applied to $\bar w$, $\bar S_n-S_n(\beta;\bar w)\pto0$, and, by the last paragraph of the proof of Lemma~\ref{lem:shift}, the same limit holds for $\Prob(\bar S_n\le x\mid\mathcal D_n)=\Prob\{\beta\in\bar R_{n,B}(\alpha)\mid\mathcal D_n\}$.

(iii) Under (A7), $\Lambda_n\dto\Lambda_\infty=\sum_j\kappa_j\zeta_j^2/(\tau-1)$. For $R_n^{\mathrm{ols}}(\alpha)$, $x>0$ is a continuity point of the distribution of $(\tau-1)\Lambda_\infty/\tau$, so (ii) yields
\[
\Prob\{\beta\in R_n^{\mathrm{ols}}(\alpha)\}\to\Prob\Big(\sum_j\kappa_j\zeta_j^2/\tau\le x\Big).
\]
As $\kappa_j/\tau\le a_j$ for every $j$, the limit is at least $\Prob(\sum_ja_j\zeta_j^2\le x)=\lim_n\Prob\{\beta\in R_n(\alpha)\}$, and as $\sum_j\kappa_j\zeta_j^2/\tau\le(\kappa_1/\tau)\sum_j\zeta_j^2$, the limit is at least $1-\alpha$ when $\kappa_1\le\tau$.

For $C_{n,\infty}$, $\Lambda^\ast>0$ is a continuity point of the distribution of $\Lambda_\infty$, and $\Psi_{x_\nu}$ is continuous and strictly decreasing with $\Psi_{x_\nu}(\Lambda^\ast)=1-\alpha$, so, by (ii), $\{\beta\in C_{n,\infty}\}=\{p_n(\beta)\ge1-\alpha\}$ differs from $\{\Lambda_n\le\Lambda^\ast\}$ by an event of probability $o(1)$, and
\[
\Prob(\beta\in C_{n,\infty})\to\Prob(\Lambda_\infty\le\Lambda^\ast)=\Prob\Big\{\sum_j\kappa_j\zeta_j^2\le(\tau-1)\Lambda^\ast\Big\}.
\]

Proposition~\ref{prop:fixedtau}(ii) applied to $\bar w$ yields $S_n(\beta;\bar w)\dto\sum_ja_j(B)\zeta_j^2$ with
\[
a_j(B)=\frac{\tau_B-1+\kappa_j}{\tau_B}=\frac{\tau-1+B\kappa_j}{\tau-1+B},
\]
and $\bar S_n$ has the same limit by the proof of (ii). With every $a_j(B)>0$, the limit has a continuous distribution, so $\Prob\{\beta\in\bar R_{n,B}(\alpha)\}=\Prob(\bar S_n\le x)\to\Prob\{\sum_ja_j(B)\zeta_j^2\le x\}$.
\end{proof}

\bibliographystyle{plainnat}
\bibliography{refs}

\begin{thebibliography}{44}
\providecommand{\natexlab}[1]{#1}
\providecommand{\url}[1]{\texttt{#1}}
\expandafter\ifx\csname urlstyle\endcsname\relax
  \providecommand{\doi}[1]{doi: #1}\else
  \providecommand{\doi}{doi: \begingroup \urlstyle{rm}\Url}\fi

\bibitem[Andrews(1991)]{Andrews1991}
Donald W.~K. Andrews.
\newblock Heteroskedasticity and autocorrelation consistent covariance matrix
  estimation.
\newblock \emph{Econometrica}, 59\penalty0 (3):\penalty0 817--858, 1991.
\newblock \doi{10.2307/2938229}.

\bibitem[Arcones(1994)]{Arcones1994}
Miguel~A. Arcones.
\newblock Limit theorems for nonlinear functionals of a stationary {G}aussian
  sequence of vectors.
\newblock \emph{The Annals of Probability}, 22\penalty0 (4):\penalty0
  2242--2274, 1994.
\newblock \doi{10.1214/aop/1176988503}.

\bibitem[Baillie et~al.(2025)Baillie, Diebold, Kapetanios, Kim, and
  Mora]{BaillieDieboldKapetaniosKimMora2025}
Richard~T. Baillie, Francis~X. Diebold, George Kapetanios, Kun~Ho Kim, and
  Aaron Mora.
\newblock On robust inference in time-series regression.
\newblock \emph{The Econometrics Journal}, 28\penalty0 (2):\penalty0 131--173,
  2025.
\newblock \doi{10.1093/ectj/utae019}.

\bibitem[Bradley(1989)]{Bradley1989}
Richard~C. Bradley.
\newblock A stationary, pairwise independent, absolutely regular sequence for
  which the central limit theorem fails.
\newblock \emph{Probability Theory and Related Fields}, 81\penalty0
  (1):\penalty0 1--10, 1989.
\newblock \doi{10.1007/BF00343735}.

\bibitem[Carta et~al.(2013)Carta, Vel{\'a}zquez, and
  Cabrera]{CartaVelazquezCabrera2013}
Jos{\'e}~A. Carta, Sergio Vel{\'a}zquez, and Pedro Cabrera.
\newblock A review of measure-correlate-predict ({MCP}) methods used to
  estimate long-term wind characteristics at a target site.
\newblock \emph{Renewable and Sustainable Energy Reviews}, 27:\penalty0
  362--400, 2013.
\newblock \doi{10.1016/j.rser.2013.07.004}.

\bibitem[Chan and Wood(1999)]{ChanWood1999}
G.~Chan and A.~T.~A. Wood.
\newblock Simulation of stationary {G}aussian vector fields.
\newblock \emph{Statistics and Computing}, 9\penalty0 (4):\penalty0 265--268,
  1999.
\newblock \doi{10.1023/A:1008903804954}.

\bibitem[Chung(2002)]{Chung2002}
Ching-Fan Chung.
\newblock Sample means, sample autocovariances, and linear regression of
  stationary multivariate long memory processes.
\newblock \emph{Econometric Theory}, 18\penalty0 (1):\penalty0 51--78, 2002.
\newblock \doi{10.1017/S0266466602181047}.

\bibitem[Cox(1958)]{Cox1958}
David~R. Cox.
\newblock Some problems connected with statistical inference.
\newblock \emph{The Annals of Mathematical Statistics}, 29\penalty0
  (2):\penalty0 357--372, 1958.
\newblock \doi{10.1214/aoms/1177706618}.

\bibitem[Dalla et~al.(2006)Dalla, Giraitis, and
  Hidalgo]{DallaGiraitisHidalgo2006}
V.~Dalla, L.~Giraitis, and J.~Hidalgo.
\newblock Consistent estimation of the memory parameter for nonlinear time
  series.
\newblock \emph{Journal of Time Series Analysis}, 27\penalty0 (2):\penalty0
  211--251, 2006.
\newblock \doi{10.1111/j.1467-9892.2005.00464.x}.

\bibitem[Davydov and Tempelman(2024)]{DavydovTempelman2024}
Youri Davydov and Arkady Tempelman.
\newblock Randomized limit theorems for stationary ergodic random processes and
  fields.
\newblock \emph{Stochastic Processes and their Applications}, 174:\penalty0
  104380, 2024.
\newblock \doi{10.1016/j.spa.2024.104380}.

\bibitem[Davydov(1970)]{Davydov1970}
Yu.~A. Davydov.
\newblock The invariance principle for stationary processes.
\newblock \emph{Theory of Probability and Its Applications}, 15\penalty0
  (3):\penalty0 487--498, 1970.
\newblock \doi{10.1137/1115050}.

\bibitem[Diebold and Inoue(2001)]{DieboldInoue2001}
Francis~X. Diebold and Atsushi Inoue.
\newblock Long memory and regime switching.
\newblock \emph{Journal of Econometrics}, 105\penalty0 (1):\penalty0 131--159,
  2001.
\newblock \doi{10.1016/S0304-4076(01)00073-2}.

\bibitem[Dudley(2002)]{Dudley2002}
R.~M. Dudley.
\newblock \emph{Real Analysis and Probability}, volume~74 of \emph{Cambridge
  Studies in Advanced Mathematics}.
\newblock Cambridge University Press, Cambridge, 2002.
\newblock \doi{10.1017/CBO9780511755347}.

\bibitem[Frisch and Waugh(1933)]{FrischWaugh1933}
Ragnar Frisch and Frederick~V. Waugh.
\newblock Partial time regressions as compared with individual trends.
\newblock \emph{Econometrica}, 1\penalty0 (4):\penalty0 387--401, 1933.
\newblock \doi{10.2307/1907330}.

\bibitem[Geyer and Meeden(2005)]{GeyerMeeden2005}
Charles~J. Geyer and Glen~D. Meeden.
\newblock Fuzzy and randomized confidence intervals and {P}-values.
\newblock \emph{Statistical Science}, 20\penalty0 (4):\penalty0 358--366, 2005.
\newblock \doi{10.1214/088342305000000340}.

\bibitem[Guo and Koul(2008)]{GuoKoul2008}
Hongwen Guo and Hira~L. Koul.
\newblock Asymptotic inference in some heteroscedastic regression models with
  long memory design and errors.
\newblock \emph{The Annals of Statistics}, 36\penalty0 (1):\penalty0 458--487,
  2008.
\newblock \doi{10.1214/009053607000000686}.

\bibitem[Haslett and Raftery(1989)]{HaslettRaftery1989}
John Haslett and Adrian~E. Raftery.
\newblock Space-time modelling with long-memory dependence: assessing
  {I}reland's wind power resource.
\newblock \emph{Journal of the Royal Statistical Society. Series C (Applied
  Statistics)}, 38\penalty0 (1):\penalty0 1--21, 1989.
\newblock \doi{10.2307/2347679}.

\bibitem[Kiefer and Vogelsang(2002)]{KieferVogelsang2002}
Nicholas~M. Kiefer and Timothy~J. Vogelsang.
\newblock Heteroskedasticity-autocorrelation robust testing using bandwidth
  equal to sample size.
\newblock \emph{Econometric Theory}, 18\penalty0 (6):\penalty0 1350--1366,
  2002.
\newblock \doi{10.1017/S026646660218604X}.

\bibitem[Kojevnikov and Song(2023)]{KojevnikovSong2023}
Denis Kojevnikov and Kyungchul Song.
\newblock Some impossibility results for inference with cluster dependence with
  large clusters.
\newblock \emph{Journal of Econometrics}, 237\penalty0 (2):\penalty0 105524,
  2023.
\newblock \doi{10.1016/j.jeconom.2023.105524}.

\bibitem[K{\"u}nsch(1989)]{Kunsch1989}
Hans~R. K{\"u}nsch.
\newblock The jackknife and the bootstrap for general stationary observations.
\newblock \emph{The Annals of Statistics}, 17\penalty0 (3):\penalty0
  1217--1241, 1989.
\newblock \doi{10.1214/aos/1176347265}.

\bibitem[Lazarus et~al.(2018)Lazarus, Lewis, Stock, and
  Watson]{LazarusLewisStockWatson2018}
Eben Lazarus, Daniel~J. Lewis, James~H. Stock, and Mark~W. Watson.
\newblock {HAR} inference: recommendations for practice.
\newblock \emph{Journal of Business \& Economic Statistics}, 36\penalty0
  (4):\penalty0 541--559, 2018.
\newblock \doi{10.1080/07350015.2018.1506926}.

\bibitem[Leung(2022)]{Leung2022}
Michael~P. Leung.
\newblock Dependence-robust inference using resampled statistics.
\newblock \emph{Journal of Applied Econometrics}, 37\penalty0 (2):\penalty0
  270--285, 2022.
\newblock \doi{10.1002/jae.2865}.

\bibitem[Longla and Peligrad(2021)]{LongPeli2021}
Martial Longla and Magda Peligrad.
\newblock New robust confidence intervals for the mean under dependence.
\newblock \emph{Journal of Statistical Planning and Inference}, 211:\penalty0
  90--106, 2021.
\newblock \doi{10.1016/j.jspi.2020.06.001}.

\bibitem[Lovell(1963)]{Lovell1963}
Michael~C. Lovell.
\newblock Seasonal adjustment of economic time series and multiple regression
  analysis.
\newblock \emph{Journal of the American Statistical Association}, 58\penalty0
  (304):\penalty0 993--1010, 1963.
\newblock \doi{10.1080/01621459.1963.10480682}.

\bibitem[Maruyama(1949)]{Maruyama1949}
Gisiro Maruyama.
\newblock The harmonic analysis of stationary stochastic processes.
\newblock \emph{Memoirs of the Faculty of Science, Kyushu University. Series A,
  Mathematics}, 4\penalty0 (1):\penalty0 45--106, 1949.
\newblock \doi{10.2206/kyushumfs.4.45}.

\bibitem[Mason and Newton(1992)]{MasonNewton1992}
David~M. Mason and Michael~A. Newton.
\newblock A rank statistics approach to the consistency of a general bootstrap.
\newblock \emph{The Annals of Statistics}, 20\penalty0 (3):\penalty0
  1611--1624, 1992.
\newblock \doi{10.1214/aos/1176348787}.

\bibitem[McElroy and Politis(2012)]{McElroyPolitis2012}
Tucker McElroy and Dimitris~N. Politis.
\newblock Fixed-b asymptotics for the studentized mean from time series with
  short, long, or negative memory.
\newblock \emph{Econometric Theory}, 28\penalty0 (2):\penalty0 471--481, 2012.
\newblock \doi{10.1017/S0266466611000405}.

\bibitem[Newey and West(1987)]{NeweyWest1987}
Whitney~K. Newey and Kenneth~D. West.
\newblock A simple, positive semi-definite, heteroskedasticity and
  autocorrelation consistent covariance matrix.
\newblock \emph{Econometrica}, 55\penalty0 (3):\penalty0 703--708, 1987.
\newblock \doi{10.2307/1913610}.

\bibitem[Pipiras and Taqqu(2017)]{PipirasTaqqu2017}
Vladas Pipiras and Murad~S. Taqqu.
\newblock \emph{Long-Range Dependence and Self-Similarity}, volume~45 of
  \emph{Cambridge Series in Statistical and Probabilistic Mathematics}.
\newblock Cambridge University Press, Cambridge, 2017.
\newblock \doi{10.1017/CBO9781139600347}.

\bibitem[Politis and Romano(1994)]{PolitisRomano1994}
Dimitris~N. Politis and Joseph~P. Romano.
\newblock The stationary bootstrap.
\newblock \emph{Journal of the American Statistical Association}, 89\penalty0
  (428):\penalty0 1303--1313, 1994.
\newblock \doi{10.1080/01621459.1994.10476870}.

\bibitem[Pr{\ae}stgaard and Wellner(1993)]{PraestgaardWellner1993}
Jens Pr{\ae}stgaard and Jon~A. Wellner.
\newblock Exchangeably weighted bootstraps of the general empirical process.
\newblock \emph{The Annals of Probability}, 21\penalty0 (4):\penalty0
  2053--2086, 1993.
\newblock \doi{10.1214/aop/1176989011}.

\bibitem[Preinerstorfer and P{\"o}tscher(2016)]{PreinerstorferPotscher2016}
David Preinerstorfer and Benedikt~M. P{\"o}tscher.
\newblock On size and power of heteroskedasticity and autocorrelation robust
  tests.
\newblock \emph{Econometric Theory}, 32\penalty0 (2):\penalty0 261--358, 2016.
\newblock \doi{10.1017/S0266466614000899}.

\bibitem[Robinson(1995)]{Robinson1995}
P.~M. Robinson.
\newblock Gaussian semiparametric estimation of long range dependence.
\newblock \emph{The Annals of Statistics}, 23\penalty0 (5):\penalty0
  1630--1661, 1995.
\newblock \doi{10.1214/aos/1176324317}.

\bibitem[Robinson and Hidalgo(1997)]{RobinsonHidalgo1997}
Peter~M. Robinson and F.~Javier Hidalgo.
\newblock Time series regression with long-range dependence.
\newblock \emph{The Annals of Statistics}, 25\penalty0 (1):\penalty0 77--104,
  1997.
\newblock \doi{10.1214/aos/1034276622}.

\bibitem[Rubin(1981)]{Rubin1981}
Donald~B. Rubin.
\newblock The {B}ayesian bootstrap.
\newblock \emph{The Annals of Statistics}, 9\penalty0 (1):\penalty0 130--134,
  1981.
\newblock \doi{10.1214/aos/1176345338}.

\bibitem[Shao(2010{\natexlab{a}})]{Shao2010}
Xiaofeng Shao.
\newblock A self-normalized approach to confidence interval construction in
  time series.
\newblock \emph{Journal of the Royal Statistical Society: Series B (Statistical
  Methodology)}, 72\penalty0 (3):\penalty0 343--366, 2010{\natexlab{a}}.
\newblock \doi{10.1111/j.1467-9868.2009.00737.x}.

\bibitem[Shao(2010{\natexlab{b}})]{Shao2010DWB}
Xiaofeng Shao.
\newblock The dependent wild bootstrap.
\newblock \emph{Journal of the American Statistical Association}, 105\penalty0
  (489):\penalty0 218--235, 2010{\natexlab{b}}.
\newblock \doi{10.1198/jasa.2009.tm08744}.

\bibitem[Smith et~al.(2011)Smith, Lott, and Vose]{SmithLottVose2011}
Adam Smith, Neal Lott, and Russ Vose.
\newblock The {I}ntegrated {S}urface {D}atabase: recent developments and
  partnerships.
\newblock \emph{Bulletin of the American Meteorological Society}, 92\penalty0
  (6):\penalty0 704--708, 2011.
\newblock \doi{10.1175/2011BAMS3015.1}.

\bibitem[Song(2016)]{Song2016}
Kyungchul Song.
\newblock Ordering-free inference from locally dependent data.
\newblock arXiv:1604.00447, 2016.

\bibitem[Taqqu(1975)]{Taqqu1975}
Murad~S. Taqqu.
\newblock Weak convergence to fractional {B}rownian motion and to the
  {R}osenblatt process.
\newblock \emph{Zeitschrift f{\"u}r Wahrscheinlichkeitstheorie und verwandte
  Gebiete}, 31\penalty0 (4):\penalty0 287--302, 1975.
\newblock \doi{10.1007/BF00532868}.

\bibitem[Tsay and Chung(2000)]{TsayChung2000}
Wen-Jen Tsay and Ching-Fan Chung.
\newblock The spurious regression of fractionally integrated processes.
\newblock \emph{Journal of Econometrics}, 96\penalty0 (1):\penalty0 155--182,
  2000.
\newblock \doi{10.1016/S0304-4076(99)00056-1}.

\bibitem[Voln{\'y}(1990)]{Volny1990}
Dalibor Voln{\'y}.
\newblock On limit theorems and category for dynamical systems.
\newblock \emph{Yokohama Mathematical Journal}, 38\penalty0 (1):\penalty0
  29--35, 1990.

\bibitem[White(1980)]{White1980}
Halbert White.
\newblock A heteroskedasticity-consistent covariance matrix estimator and a
  direct test for heteroskedasticity.
\newblock \emph{Econometrica}, 48\penalty0 (4):\penalty0 817--838, 1980.
\newblock \doi{10.2307/1912934}.

\bibitem[Zheng(1987)]{Zheng1987}
Zhongguo Zheng.
\newblock Random weighting method.
\newblock \emph{Acta Mathematicae Applicatae Sinica}, 10\penalty0 (2):\penalty0
  247--253, 1987.
\newblock In Chinese.

\end{thebibliography}

\end{document}